\documentclass[twocolumn]{aastex631}

\usepackage{amsmath}        
\usepackage{gensymb}        

\begin{document}

\title{Superluminal proper motion in the X-ray jet of Centaurus A}

\author[0000-0002-5924-4822]{David Bogensberger}
\affiliation{Department of Astronomy, The University of Michigan, 1085 South University Avenue, Ann Arbor, Michigan, 48103, USA}

\author[0000-0003-2869-7682]{Jon M Miller}
\affiliation{Department of Astronomy, The University of Michigan, 1085 South University Avenue, Ann Arbor, Michigan, 48103, USA}

\author{Richard Mushotzky}
\affiliation{Department of Astronomy, University of Maryland, College Park, MD 20742, USA}

\author[0000-0002-0167-2453
]{W. N. Brandt}
\affiliation{Department of Astronomy and Astrophysics, 525 Davey Lab, The Pennsylvania State University, University Park, PA 16802, USA}
\affiliation{Institute for Gravitation and the Cosmos, The Pennsylvania State University, University Park, PA 16802, USA}
\affiliation{Department of Physics, 104 Davey Laboratory, The Pennsylvania State University, University Park, PA 16802, USA}

\author[0000-0002-0273-218X]{Elias Kammoun}
\affiliation{IRAP, Universit\'{e} de Toulouse, CNRS, UPS, CNES, 9, Avenue du Colonel Roche, BP 44346, F-31028, Toulouse Cedex 4, France}

\affiliation{INAF -- Osservatorio Astrofisico di Arcetri, Largo Enrico Fermi 5, I-50125 Firenze, Italy
}

\affiliation{Dipartimento di Matematica e Fisica, Universit\`{a} Roma Tre, via della Vasca Navale 84, I-00146 Rome, Italy}

\author{Abderahmen Zoghbi}
\affiliation{Department of Astronomy, University of Maryland, College Park, MD 20742}
\affiliation{HEASARC, Code 6601, NASA/GSFC, Greenbelt, MD 20771}
\affiliation{CRESST II, NASA Goddard Space Flight Center, Greenbelt, MD 20771}

\author{Ehud Behar}
\affiliation{Faculty of Physics, Technion - Israel Institute of Technology, Haifa, 3200003, Israel}



\begin{abstract}
\noindent
The structure of the jet in Cen A is likely better revealed in X-rays than in the radio band, which is usually used to investigate jet proper motions. In this paper, we analyze Chandra ACIS observations of Cen A from 2000 to 2022 and develop an algorithm for systematically fitting the proper motions of its X-ray jet knots. Most of the knots had an apparent proper motion below the detection limit. However, one knot at a transverse distance of $520~\mathrm{pc}$ had an apparent superluminal proper motion of $2.7\pm0.4~\mathrm{c}$. This constrains the inclination of the jet to be $i<41\pm6\degree$, and the velocity of this knot to be $\beta>0.94\pm0.02$. This agrees well with the inclination measured in the inner jet by the EHT, but contradicts previous estimates based on jet and counterjet brightness. It also disagrees with the proper motion of the corresponding radio knot, of $0.8\pm0.1~\mathrm{c}$, which further indicates that the X-ray and radio bands trace distinct structures in the jet. There are four prominent X-ray jet knots closer to the nucleus, but only one of these is inconsistent with being stationary. A few jet knots also have a significant proper motion component in the non-radial direction. This component is typically larger closer to the center of the jet. We also detect brightness and morphology variations at a transverse distance of $100~\mathrm{pc}$ from the nucleus.

\end{abstract}



\section{Introduction} \label{sec:intro}

Some active galactic nuclei (AGNs) launch powerful relativistic jets that extend over Mpc distances \citep{2020NewAR..8801539H, 2022A&A...660A...2O}. Jets produce very prominent radio emission \citep{2019ARA&A..57..467B}, but have also been observed to be unusually and unexpectedly bright in X-rays \citep{2006ARA&A..44..463H} and gamma-rays \citep{2016ARA&A..54..725M}, which sparked an ongoing debate about the physical origin of the emission process. 

A single synchrotron emission spectrum cannot account for both the radio and X-ray emission \citep{2016ApJ...826..109C, 2019ARA&A..57..467B}. Instead, one proposed model invoked the up-scattering of microwave photons from the cosmic microwave background (CMB) by relativistic electrons through the inverse Compton (IC) effect \citep{2001MNRAS.321L...1C, 2006ApJ...652..163M, 2011ApJ...730...92H}. However, various studies of X-ray jets have challenged this IC-CMB model. For instance, variability in the brightness of a significant fraction of X-ray jets has been detected \citep{2010ApJ...714L.213M, 2016MNRAS.455.3526H, 2023NatAs...7..967M}, in stark contrast to the expectation of consistent emission on $\sim10^6~\mathrm{yr}$ timescales \citep{2007RMxAC..27..188H}. Other potential flaws in the model include the requirements of bulk deceleration, and the inconsistencies with observed velocity structures \citep{2006MNRAS.366.1465H}. Observations of high-redshift, kpc-scale jets also contradict the IC-CMB model predictions \citep{2004AJ....128..523B, 2006AJ....131.1914L}.

The IC-CMB model may, however, still be an important mechanism of X-ray emission in high redshift jets \citep{2006MNRAS.366.1465H, 2023NatAs...7..967M, 2024MNRAS.530..360Z}. High-redshift jets with IC-CMB emission are sometimes only detected in X-rays, not in the radio band \citep{2016ApJ...816L..15S}, or have a very steep radio spectrum \citep{2020MNRAS.497..988W}.

The X-ray emission in low redshift jets is possibly instead dominated by a second synchrotron emission component, other than the one in the radio band \citep{2023NatAs...7..967M}. This X-ray emission must originate from regions much smaller than the width of the jet, to explain the observed brightness variation on timescales of months \citep{2010ApJ...714L.213M, 2016MNRAS.455.3526H}. For Cen A, this is regarded as the most likely cause of the observed X-ray emission in the jet \citep{2020Natur.582..356H}.

Regardless of the exact emission mechanism, particles within the jet may travel at relativistic speeds. At particular inclinations to the line of sight, this can result in apparent superluminal proper motions. The reason for this is that the distance from a region of the jet to the observer may decrease at a rate of a significant fraction of the speed of light.  The equation for the apparent proper motion ($\beta_{\mathrm{app}}c$, where $c$ is the speed of light) as a function of the inclination ($i$), and the actual speed ($\beta c$) is: 

\begin{equation} \label{eq:propmot}   
    \beta_{\mathrm{app}} = \frac{\beta \sin{i}}{1-\beta \cos{i}}.
\end{equation}

\noindent
If the inclination is not known, a measurement of a superluminal proper motion ($\beta_{\mathrm{app}}>1$) can provide an upper limit on the inclination: 

\begin{equation} \label{eq:maxinc}
    i_{\mathrm{max}} = \cos^{-1}\left(\frac{\beta_{\mathrm{app}}-1}{\beta_{\mathrm{app}}+1}\right).
\end{equation}

\noindent
It also provides a lower limit on the speed of the jet:
\begin{equation} \label{eq:minbeta}
    \beta_{\mathrm{min}} = \frac{\beta_{\mathrm{app}}}{\sqrt{1+\beta_{\mathrm{app}}^2}}.
\end{equation}

\noindent
In the following, we describe the proper motions as fractions of the speed of light. Therefore, they are expressed without any units. 

Jets are not uniform but contain extended features of enhanced brightness. These are referred to as jet knots and are seen in both the radio and X-ray bands. In some cases, there is a good agreement between the radio and the X-ray knots \citep{2015ApJS..220....5M}, but in other jets, there are systematic offsets between them \citep{2023ApJS..265....8R}. Jet knots can be regarded as blobs that move along with the jet, which have proper motions equivalent to that of individual particles in the jet \citep{1999MNRAS.309..233O, 2003ApJ...592..332V}. Alternatively, they have also been regarded as shocks or standing waves that move through the jet medium, and therefore do not represent the motion of individual particles. This could be due to magnetic recompression \citep{2002LNP...589...41S}, the Kelvin-Helmholtz instability \citep{2021A&A...649A.150B}, or soliton waves \citep{2005ApJ...625...37L}. Some jet knots have also been interpreted as the shock produced by the interaction of the jet with an obstacle, such as a gas cloud, or a high mass-loss star \citep{2003ApJ...593..169H, 2007ApJ...655..769C, 2015MNRAS.447.1001W}.  

Measurements of jet knot proper motions have been performed in many systems in the radio band, with apparent proper motions of up to $\beta_{\mathrm{app}}\approx 46$ for the blazar PKS 1510-089 \citep{2005AJ....130.1418J}. In X-rays, only a very limited number of apparent proper motion measurements have been performed, due to the constraints on X-ray angular resolution. \emph{Chandra} has the best angular resolution of any current X-ray telescope, of $0.492"$, but that is only sufficient to resolve apparent proper motion in the closest AGN jets. In M87, two X-ray jet knots had detectable apparent superluminal proper motions of $6.3 \pm 0.4$, and $2.4 \pm 0.6$ \citep{2019ApJ...879....8S, 2024arXiv240419272T}. These results agree with the radio measurements of the same jet knots \citep{1995ApJ...447..582B}. 

The galaxy Centaurus A (Cen A, hereafter) hosts the nearest AGN. It also launches a jet and is classified as an FR Type I radio galaxy. Its distance of $3.8 \pm 0.1 ~ \mathrm{Mpc}$ \citep{2010PASA...27..457H} allows us to resolve smaller structures in the jet, than is possible for any other radio galaxy. Therefore, Cen A is an ideal source for studying jet morphology, kinematics, and evolution. The jet is prominently seen in both the radio and X-ray bands, but not in the optical \citep{1996ApJ...459..535S}. 

The jet is inclined, with the northeastern side pointing toward us. The counterjet points to the southwest and is significantly fainter. The inclination of the jet relative to the line of sight is, however, not well constrained. \citet{1979AJ.....84..284D} investigated the distribution of OB stars and H II regions in the galaxy. Assuming a circular shape, they estimate the plane of the galaxy to be inclined by $72\pm3\degree$. Jets are, however, often misaligned with their host galaxy \citep{2012MNRAS.425.1121H}. \citet{1994ApJ...426L..23S} studied the hard X-ray and gamma-ray spectrum of the Cen A jet. By assuming the jet emission to be produced by Compton scattered photons emitted by the nucleus into the line of sight, they estimated the jet to be inclined by $61\pm5\degree$. \citet{1996ApJ...466L..63J}, and \citet{1998AJ....115..960T} performed a similar calculation for the radio band, and found an inclination of $60-77\degree$, and $50-80\degree$, respectively. \citet{2003ApJ...593..169H} repeated the above analysis for the jet and counterjet knots closest to the nucleus, and also used estimates of the bulk speed. This resulted in an inclination estimate of $\approx 15\degree$. \citet{2014A&A...569A.115M} studied the radio structure closest to the nucleus on sub-pc scales, and used it to estimate an inclination of $12-45\degree$. Finally, EHT observations on mpc scales also found the inclination to be $12–45\degree$, under the assumption that the jet does not bend \citep{2021NatAs...5.1017J}.

The Cen A jet has been detected as close as $11~\mathrm{AU}$ to the SMBH powering it \citep{2021NatAs...5.1017J}, and extends for $4~\mathrm{kpc}$ \citep{2002ApJ...569...54K}, beyond which it expands into a plume. Radio lobes from the jet have been found to extend for up to $600~\mathrm{kpc}$ from the nucleus \citep{2013ApJ...766...48S}. These are all projected distances, perpendicular to the line of sight. The actual distances involved are found by dividing by $\sin{i}$, where $i$ is the inclination angle. 

The jet appears very different in the X-ray and radio bands ranges, with some knots being "X-ray-only", and others being "radio-only"  \citep{2003ApJ...593..169H, 2010ApJ...708..675G}. Whereas the X-ray jet consists of discrete, well-separated bright knots, the radio band shows a more uniform, diffuse emission within the inner kpc \citep{2003ApJ...593..169H}. There also appears to be an offset between the location of some X-ray and radio knots \citep{2003ApJ...593..169H}. These differences are crucial for unraveling the different emission mechanisms at opposite ends of the spectrum, and warrant further investigation.  

\citet{1998AJ....115..960T} performed very-long-baseline interferometry to study the proper motions in Cen A. They found that the brightest jet knots moved with an apparent proper motion of $\approx 0.1$. However, they also interpreted the detected variability in the jet knots as evidence of a much faster jet flow, with a proper motion of $\approx 0.45$.

\citet{2003ApJ...593..169H} measured the proper motion of a greater number of radio jet knots in Cen A and found them to be consistently moving with an apparent proper motion of $\approx 0.5$, if the stationary jet knots studied by \citet{1998AJ....115..960T} were disregarded. \citet{2003ApJ...593..169H} interpreted this proper motion as being indicative of the bulk flow velocity of the jet, when corrected for the inclination. 

\citet{2010ApJ...708..675G} performed a more detailed investigation into the radio proper motions of the Cen A jet. Most of the jet knots were found to have inconclusive proper motions, or ones consistent with being stationary. However, three knots had non-zero proper motions, of $0.534^{+0.06}_{-0.02}$, $0.338^{+0.22}_{-0.15}$, and $0.802^{+0.15}_{-0.09}$.

\citet{2019ApJ...871..248S} investigated the X-ray variability and proper motion of the jet in Cen A in five observations observations taken in 2002, 2003, 2009, and 2017. They found that the three brightest jet knots had no detectable proper motion, with upper limits of $0.1$. After excluding these three jet knots, they cross-correlated images of the jets in different epochs and thereby estimated a speed for the rest of the X-ray jet to be $\approx 0.68\pm0.20$.

In this paper, we study the proper motion, and variability of the Cen A jet throughout the entire archive of \emph{Chandra} observations. We measure the proper motions of individual jet knots over a longer interval and analyze their direction of motion. This paper is structured as follows. Section \ref{sec:obs} describes the selection of observations and their properties. Section \ref{sec:datanalys} outlines the data analysis procedure we used. In Section \ref{sec:align}, we discuss the methods we used to align the different observations. Candidate jet knots, and jet variability close to the nucleus are investigated in Section \ref{sec:newknots}. Following that, Section \ref{sec:jetvelocity} discusses the proper motions measured for the jet knots. It details the methodology, and the error analysis, and then presents the results. In Section \ref{sec:discussion}, we discuss our findings and compare them to those of previous works. Finally, we conclude our main findings in Section \ref{sec:conc}

\begin{figure}[h]
\resizebox{\hsize}{!}{\includegraphics{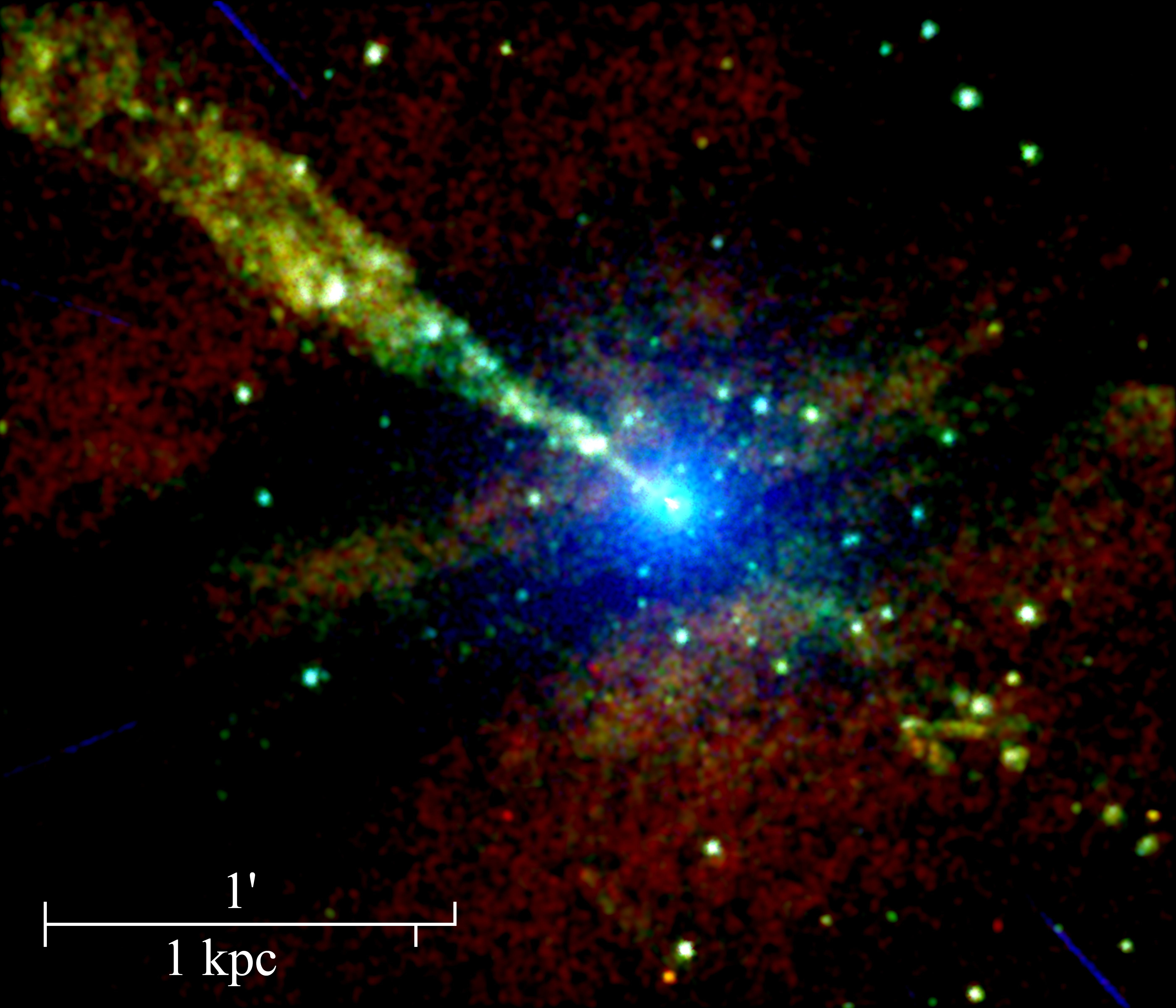}}
\caption{Adaptively smoothed, logarithmically scaled, background-subtracted false-color image of the combined \emph{Chandra} observations of Cen A used in this work. The red, green, and blue colors represent the three energy bands: $0.3-1.0 ~\mathrm{keV}$, $1.0-3.0 ~\mathrm{keV}$, and $3.0-10.0 ~\mathrm{keV}$, respectively. The individual images were aligned according to the method described in Section \ref{sec:align}. Two blue lines are the first-order HETGS grating arms from one of the observations. In this, and the following images, the brightness is scaled by the number of counts per pixel in the merged image. The field of view of this image is $2.87'\times2.46'$, which corresponds to $3.17~\mathrm{kpc}\times2.72~\mathrm{kpc}$ at $3.8~\mathrm{Mpc}$. The nucleus is located at $\mathrm{RA}=13\mathrm{h}\, 25\mathrm{m}\, 27.6\mathrm{s}$, $\mathrm{Dec}=-43\degree\, 01'\, 08.8 ''$.
 \label{fig:RGB}}
\end{figure}

\section{Observations} \label{sec:obs}

Cen A has frequently been observed by the \emph{Chandra X-ray Observatory} \citep[\textit{Chandra};][]{2000SPIE.4012....2W} since its launch. In this paper, we focus on the 63 observations taken with the Advanced CCD Imaging Spectrometer \citep[ACIS;][]{2003SPIE.4851...28G}, with data in the public archive at the time of writing.

Out of those observations, we discarded the ones in which the central source was further than $4'$ away from the center of the image. At large off-axis angles, \emph{Chandra} images suffer from spherical and other aberrations and thus are too distorted to use in our analysis.

\emph{Chandra} images also feature readout streaks. Several observations were performed with roll angles that placed the streak within the angular width of the jet. To avoid the streak, and its removal from having an impact on our study of the Cen A jet, we discarded all such observations. The remaining set of 34 observations, which were used throughout the rest of the paper, are listed in Table \ref{tab:Cobs}, and are contained in the Chandra Data Collection (CDC) 251 (\dataset[doi:10.25574/cdc.251]{https://doi.org/10.25574/cdc.251}). Figure \ref{fig:RGB} displays the false-color image of this data set. 

\begin{figure}[h]
\resizebox{\hsize}{!}{\includegraphics{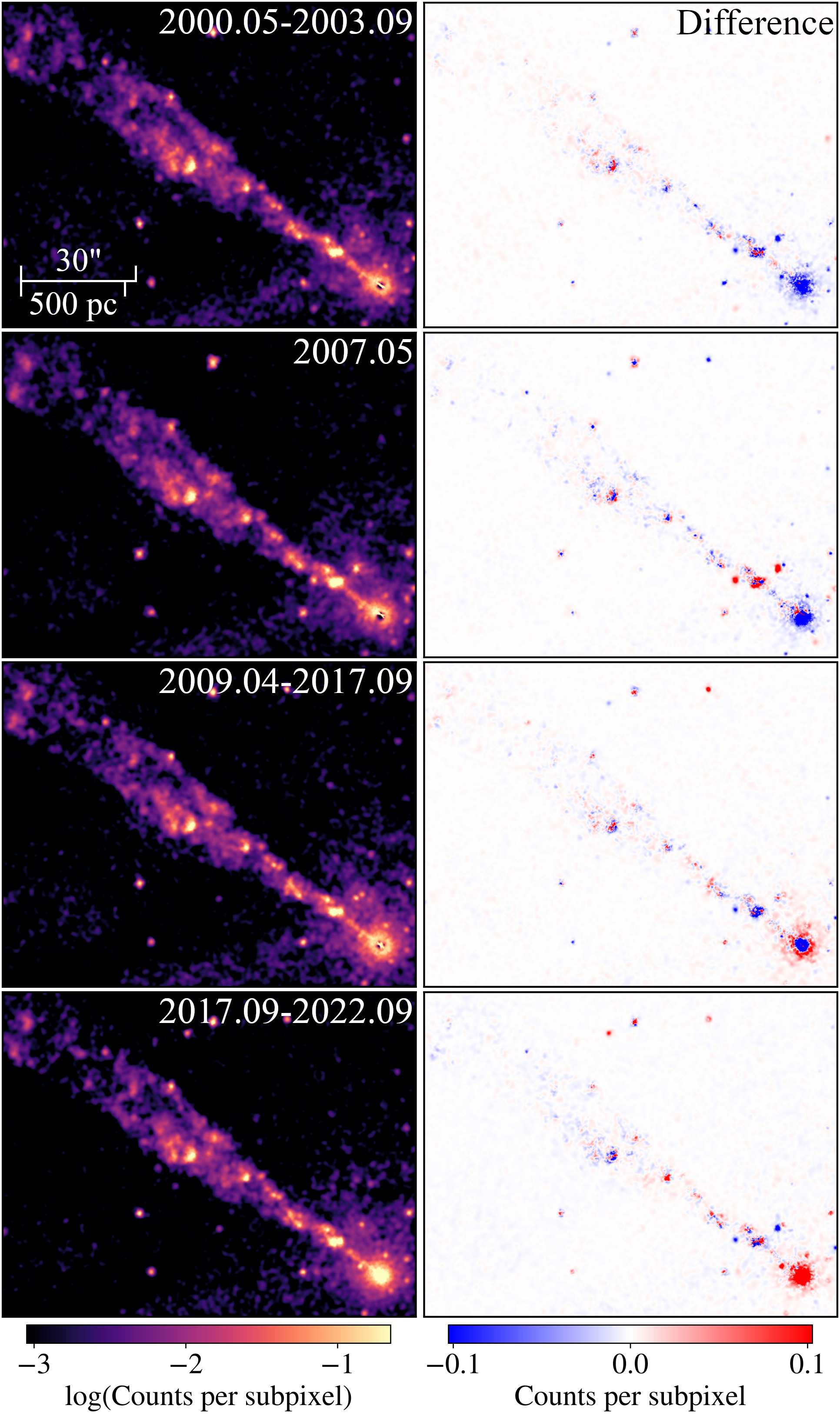}}
\caption{Images of the jet in the four groups of observations. On the left are adaptively-smoothed, logarithmically scaled, background-subtracted images of the jet in the $0.8-3.0~\mathrm{keV}$ energy band. On the right are the difference images of each group, compared to the total image. The two images being compared were scaled by the number of background-subtracted counts in the jet. Red depicts regions that are brighter in that group of observations than in the rest of the observations. Blue shows regions that are fainter in that group. The pileup at the nucleus generates weird structures, including an apparent hole, but these are not real. The field of view of this figure is $1.79'\times1.41'$, corresponding to $1.98~\mathrm{kpc}\times1.56~\mathrm{kpc}$. A colorbar is included for the fourth group. The colorbars for the other panels are similar but with a range adjusted by the number of background-subtracted counts in the jet.
 \label{fig:Jetimgdiff}}
\end{figure}

We divided these observations into four distinct groups, to visualize changes that occurred throughout these 22 years of observations. These groups were selected to contain a comparable number of background-subtracted source counts in the jet. The four groups are distinguished by horizontal lines in Table \ref{tab:Cobs}. Figure \ref{fig:Jetimgdiff} shows the jet in these four groups of observations, and the difference between that group and the total merged image. 

\begin{table}[h]
\centering
\setlength{\tabcolsep}{3pt}
\def\arraystretch{1.0}
\begin{tabular}{c|c|c|c|c|c}
    \textbf{ObsID} & \textbf{Date} & \textbf{Instrument} & \textbf{Offset} & $\boldsymbol{T ~ (\mathrm{ks})}$ & $\boldsymbol{C}$\\ \hline
    962 & 2000-05-17 & ACIS-I & $11.1''$ & 36.50 & 3371 \\
    2978 & 2002-09-03 & ACIS-S & $24.7''$ & 44.59 & 5454 \\
    3965 & 2003-09-14 & ACIS-S & $24.8''$ & 49.52 & 6040 \\ \hline
    8489 & 2007-05-08 & ACIS-I & $72.8''$ & 93.94 & 8883 \\
    8490 & 2007-05-30 & ACIS-I & $141.7''$ & 94.43 & 8594 \\ \hline
    10725 & 2009-04-26 & ACIS-I & $165.6''$ & 4.97 & 506 \\
    10722 & 2009-09-08 & ACIS-S & $11.0''$ & 49.40 & 6701 \\
    11846 & 2010-04-26 & ACIS-I & $211.8''$ & 4.69 & 444 \\
    12156 & 2011-06-22 & ACIS-I & $109.1''$ & 4.99 & 468 \\
    13303 & 2012-04-14 & ACIS-I & $200.6''$ & 5.51 & 478 \\
    15294 & 2013-04-05 & ACIS-I & $4.5''$ & 5.02 & 310 \\
    16276 & 2014-04-24 & ACIS-I & $190.3''$ & 5.02 & 413 \\
    17890 & 2016-02-24 & ACIS-S & $29.6''$ & 9.93 & 928 \\
    17891 & 2016-03-22 & ACIS-S & $29.6''$ & 9.92 & 922 \\
    18461 & 2016-04-23 & ACIS-I & $92.1''$ & 5.12 & 437 \\
    19747 & 2017-05-15 & ACIS-I & $11.0''$ & 5.00 & 346 \\
    19521 & 2017-09-17 & ACIS-S & $11.0''$ & 14.87 & 1404 \\ \hline
    20794 & 2017-09-19 & ACIS-S & $11.0''$ & 107.24 & 10008 \\
    21698 & 2019-04-29 & ACIS-I & $158.8''$ & 5.01 & 346 \\
    22714 & 2020-05-03 & ACIS-I & $175.3''$ & 4.89 & 327 \\
    24322 & 2022-01-25 & HETG & $11.2''$ & 25.32 & 617 \\ 
    23823 & 2022-04-04 & HETG & $11.2''$ & 27.81 & 730 \\ 
    24321 & 2022-05-04 & HETG & $11.2''$ & 13.95 & 276 \\ 
    26405 & 2022-05-04 & HETG & $11.2''$ & 13.95 & 286 \\ 
    24319 & 2022-06-02 & HETG & $11.2''$ & 28.27 & 656 \\ 
    24325 & 2022-07-09 & HETG & $11.2''$ & 29.66 & 762 \\ 
    24323 & 2022-07-11 & HETG & $11.2''$ & 20.51 & 485 \\ 
    26453 & 2022-07-12 & HETG & $11.2''$ & 9.33 & 223 \\ 
    24318 & 2022-07-19 & HETG & $11.2''$ & 28.79 & 742 \\
    24324 & 2022-08-10 & HETG & $11.2''$ & 29.66 & 809 \\ 
    24320 & 2022-09-07 & HETG & $11.2''$ & 13.02 & 351 \\ 
    24326 & 2022-09-07 & HETG & $11.2''$ & 13.02 & 359 \\ 
    27344 & 2022-09-08 & HETG & $11.2''$ & 13.02 & 343 \\ 
    27345 & 2022-09-09 & HETG & $11.2''$ & 13.02 & 323 \\ \hline
\end{tabular}
\caption{Properties of the \emph{Chandra} observations of Cen A used in this paper. $T$ denotes the total exposure time. $C$ denotes the total number of counts detected within the jet sheath in each observation, in the $0.2-10~\mathrm{keV}$ range. The selected region includes the two jet knots AX1 and AX2, but excludes everything closer to the nucleus, as that is contaminated from the nuclear emission. The detected counts are not background subtracted. Horizontal lines distinguish groups of observations that have a comparable number of counts in the jet, which are used to showcase variability. HETG denotes the instrument configuration of ACIS-S with the HETG. 
\label{tab:Cobs}}
\end{table}

\section{Data analysis} \label{sec:datanalys}

We generated reprocessed event files using the command \texttt{chandra\_repro} from the Chandra Interactive Analysis of Observations software \citep[CIAO;][]{2006SPIE.6270E..1VF} version 4.15. The readout streak in the observations was removed with the task \texttt{acisreadcorr}, with a streak width (\texttt{dx}) of 4 pixels, and a streak avoidance region (\texttt{dy}) of 50 pixels. 

Subpixel images at a scale of $1/16$ pixels were created for different energy bands, using the CIAO task \texttt{dmcopy}, and the energy-dependent sub-pixel event-repositioning \citep[EDSER;][]{2004ApJ...610.1204L} algorithm. We selected this subpixel size in order to be able to shift images and align them relative to each other with a comparable precision. We selected a region of $300\times350$ native pixels to investigate. Each native pixel has a size of $0.492''$, corresponding to a perpendicular distance of $9.1 ~\mathrm{pc}$ at a distance of $3.8~\mathrm{Mpc}$. The location of this region within the image was selected based on an initial estimate of the location of the central source, which we aimed to place at the native pixel [200, 150] in the selected region. The selected region is depicted in Fig. \ref{fig:RGB}.

We also created exposure-corrected images using the CIAO task \texttt{fluximage}. We specified \texttt{bands=broad}, and selected the region of interest, as described above.

\section{Alignment of images} \label{sec:align}

\begin{figure}[h]
\resizebox{\hsize}{!}
{\includegraphics{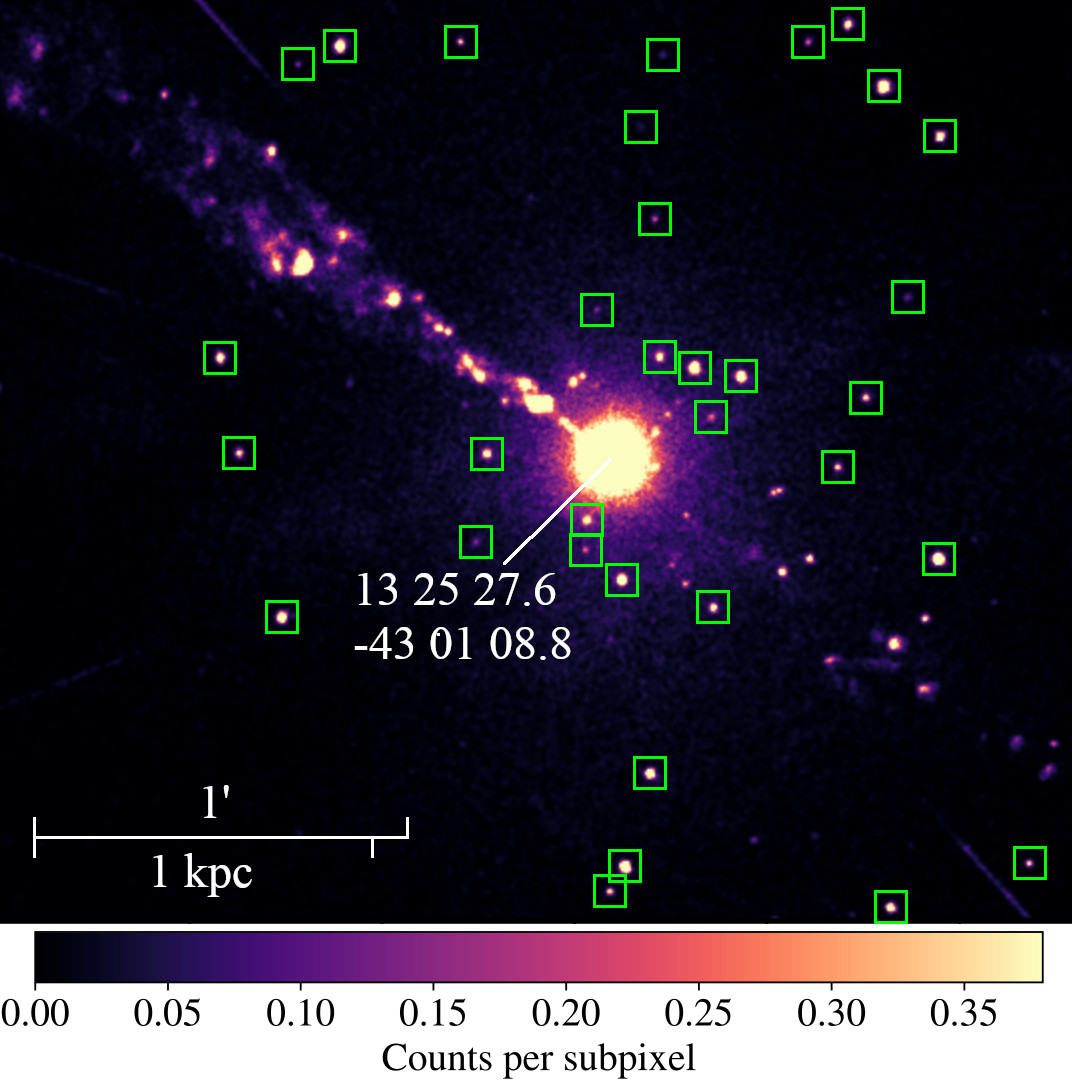}}
\caption{The 35 selected aligning point sources, and the regions around them, that are used in the optimal alignment of the 34 images. This image depicts the same region as in Fig. \ref{fig:RGB}, for the $0.2-10~\mathrm{keV}$ energy band. 
 \label{fig:alignsrc}}
\end{figure}

For aligning the images, we identified a set of 35 bright point sources in the total image, shown in Fig. \ref{fig:alignsrc}. These point sources were selected based on the condition that they had sufficient counts in the total merged image, for their point spread function (PSF) to be approximately fitted by a two-dimensional Gaussian function, with at least 50 source counts. These point sources were selected to be distinct from the jet, counterjet, and nucleus, as well as be surrounded by a stable, consistent background without large gradients. We also excluded point sources that are located within $1.5'$ of other sources, as those could interfere with the efforts to align the images. This was due to our selection of the size of the region around each point source. In the following, we assume that these sources have no significant proper motion between the different observations. 

We started with an initial visual estimate of the offset between the different observations. We fit for the center of the brightest three aligning sources in individual images, to improve on the initial estimate. Next, we selected and extracted $161\times161$ subpixel regions centered around each of the selected sources for each observation. These are shown as green boxes in Fig. \ref{fig:alignsrc}. The centers of the aligning sources were determined by fitting a two-dimensional Gaussian function to the distribution of counts created by combining all observations with the previous alignment. These regions were subsequently placed next to each other in a $6\times6$ grid, with the last position left out.

This grid of images of the aligning sources from observation $i$ was subsequently cross-correlated against a similar grid created by combining all observations, except $i$, using the previous estimate for the offset between the different images. We found the optimal shift of observation $i$ relative to the previous offset estimate from the peak of the cross-correlation of these two image grids. We subsequently updated the offset for observation $i$ accordingly, before repeating the above for observation $i+1$. This procedure was applied to all observations and was then repeated several times, to successively improve the alignment of the subpixel images. 

Some of the observations were performed with a 1/8 window size, which therefore excluded many of the aligning sources. Some of these observations also had short exposures, and were afflicted by the change in the ACIS effective area. This presented a challenge for the alignment of images via cross-correlation, as it depends on the few subpixels with non-zero counts in the selected regions around a small number of aligning sources. To counteract this issue, we also investigated improvements to the alignment of images by cross-correlating smoothed images of the aligning sources. For this purpose, we used a Gaussian smoothing algorithm, with $\sigma=8$ subpixels. A subpixel here again has a size of $1/16$ native pixels. 

As Fig. \ref{fig:alignsrc} shows, there are more aligning sources located towards the northwest than there are in the southeast corner of the image. This means that the aligned images could be somewhat biased towards better alignment of the northwest part. This might be an issue due to the different pointing centers of the observations, and the effect that varying off-axis angles have on the PSFs of the aligning sources. To counteract this, we also investigated alignments that were based on the cross-correlations of a selection of point sources that corresponded to a more uniform distribution across the image. 

These varied methods yielded different corrections to the initial alignment of observations. The relative alignment of the longest exposure observations was almost identical between the different methods, but other observations had shifts that differed by a few subpixels. We compared the accuracy of the different alignments, by calculating the mean standard deviation of the best-fit two-dimensional Gaussian function to the PSF of all 35 aligning sources. We selected the alignment that yielded the lowest value. The average absolute value of the shift of the images during this alignment process, relative to the initial estimate, was 6 subpixels.

To quantify the accuracy of this alignment, we also compared the standard deviations of these best fits to the distribution of counts in the aligning sources, for individual observations, and in the total merged image. We only considered instances in which at least 20 background-subtracted source counts were fitted for the aligning source. On average, the standard deviation of the PSF of aligning sources in individual images was found to be $0.300"$. For the merged image, it was found to be $0.305"$. Therefore, we estimate the alignment of images to be accurate to within a standard deviation of $0.05"$. This is significantly smaller than the size of the PSF, as the location of its center can be determined with greater precision.

This may underestimate the actual deviation in the positions of equivalent sources in different observations since this estimate is based on the same sources that were used to align the images. We chose against using half of the sample to align the images, and the second half to test the alignment, since we aimed to achieve the most accurate alignment possible. 

We did not consider a rotation or scaling of the images, as we found this alignment to already be sufficiently accurate for this work. The short exposure times, varying pointing centers, and different window sizes of some observations present further challenges for a more in-depth analysis of the image alignment. 

\begin{figure}[h]
\resizebox{\hsize}{!}
{\includegraphics{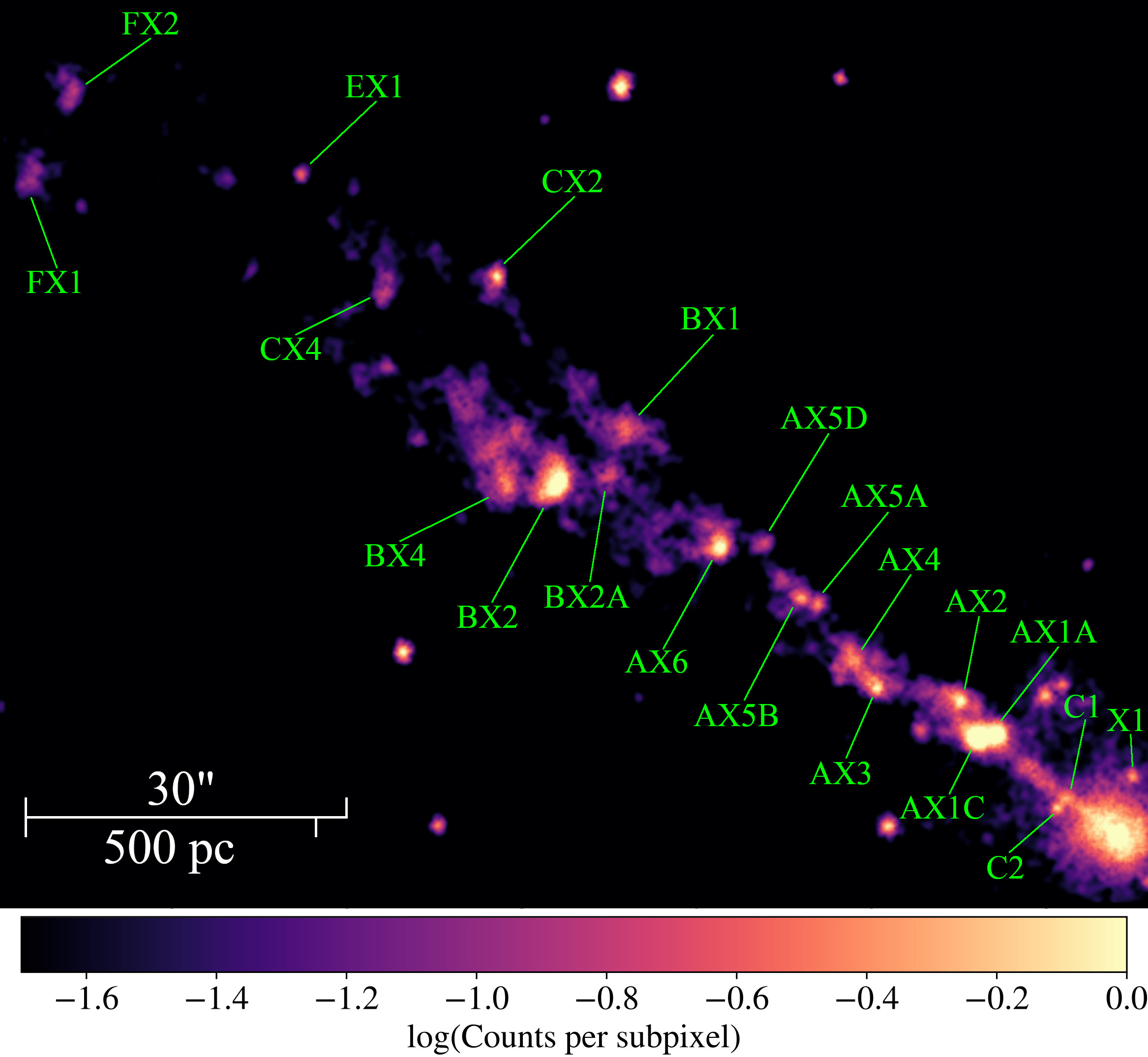}}
\caption{The X-ray jet, with labeled jet knots. The image is of the $0.8-3.0~\mathrm{keV}$ energy band and has been adaptively smoothed, logarithmically scaled, and background subtracted. It is the merged image from all \emph{Chandra} observations of Cen A used in this work. The labels of the individual jet knots are adapted from those used by \citet{2007ApJ...670L..81H, 2019ApJ...871..248S}, but with slight modifications. The two potential new candidate jet knots have been labeled as C1, and C2. X1 refers to a possible X-ray binary located at a similar distance from the nucleus as C1 and C2, and is compared with them in Section \ref{sec:newknots}. The depicted region is identical to that of Fig. \ref{fig:Jetimgdiff}
 \label{fig:jetknotlabel}}
\end{figure}

\begin{figure}[h]
\resizebox{\hsize}{!}{\includegraphics{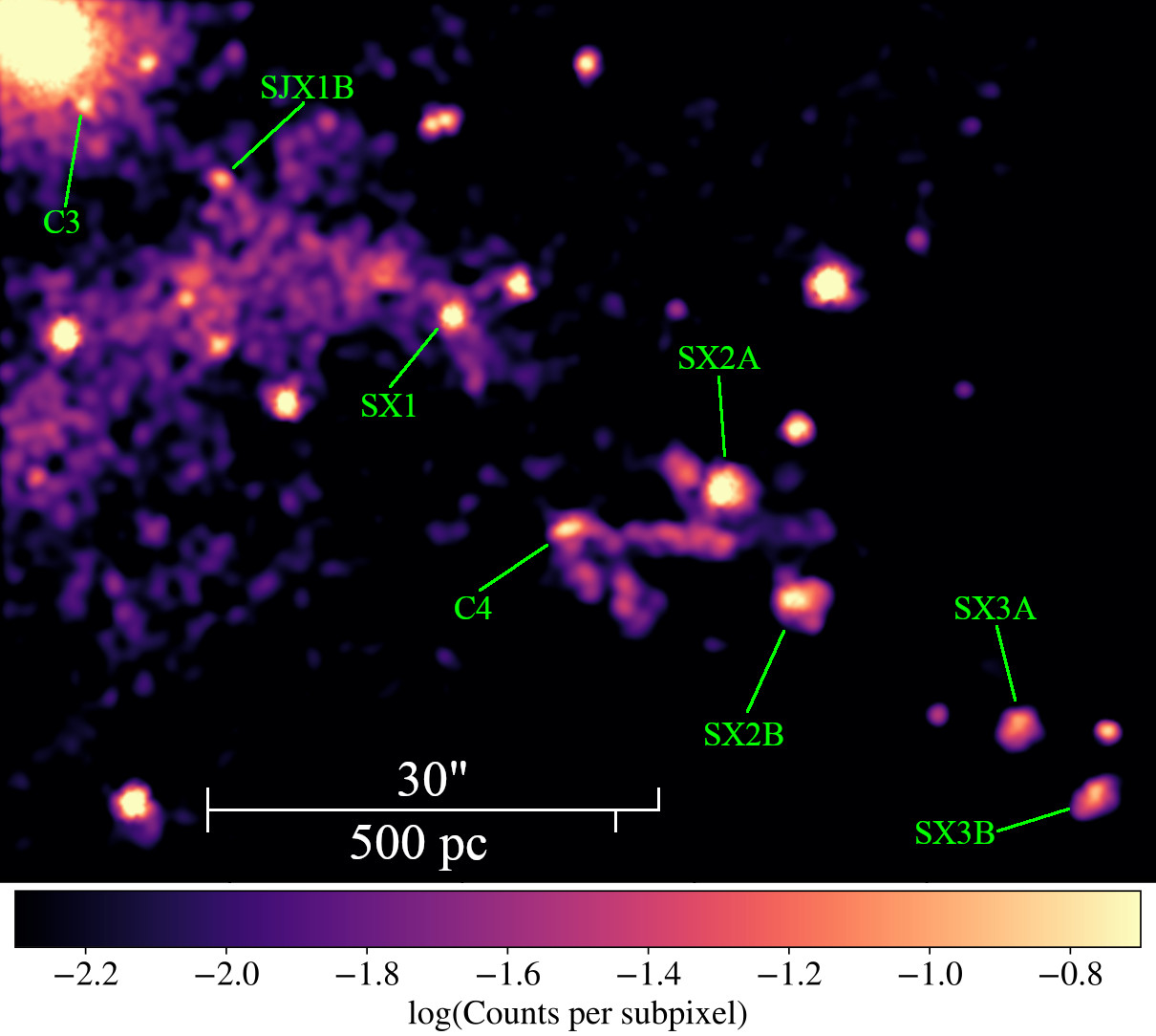}}
\caption{The X-ray counterjet, with labeled counterjet knots. C3 is a point-like source somewhat near to the counterjet axis, which is investigated along with C1, C2, and X1 in Section \ref{sec:newknots}. C4 is a point-like source at a large angle to the counterjet axis but with a curious `V'-like structure of emission behind it. The field of view of this figure is $1.28'\times0.97'$, which corresponds to $1.42~\mathrm{kpc}\times1.08~\mathrm{kpc}$. More detail is provided in Fig. \ref{fig:jetknotlabel}.
 \label{fig:cjetknotlabel}}
\end{figure}

Figures \ref{fig:jetknotlabel} and \ref{fig:cjetknotlabel} label the jet, and counter jet knots that are analyzed in more detail in the next sections. These labels are based on those used by \citet{2002ApJ...569...54K, 2003ApJ...593..169H, 2010ApJ...708..675G}. We excluded several faint knots and ones that have no unambiguously identifiable shape or center. A few bright point sources within the angular width of the jet were only present in some observations, but not detectable in others. These are likely to be X-ray binaries and were subsequently not investigated in more detail. For instance, the source that was labeled as AX2A by \citet{2010ApJ...708..675G} was only detectable in observations 8489 and 8490, and is likely to not be a jet knot.

Many of the jet knots are extended, with a larger PSF than point-like sources \citep{2002ApJ...569...54K, 2007ApJ...670L..81H}. For example, the best-fitted standard deviation of the distribution of counts in jet knot BX2 in the vertical direction is $0.62''$. In contrast, the aligning sources have a PSF with an average standard deviation of $0.28''$. Sources lying within the jet stream, which have an extended PSF can thus be identified as jet knots, and distinguished from XRBs or background AGNs. However, several jet knots, such as EX1, have a PSF that cannot be distinguished from that of a point-like source. For those sources, other methods are needed to distinguish between jet knots and point-like sources.

\citet{2010ApJ...708..675G} further investigated the X-ray spectra of jet knots and found that they are described by a very shallow power-law, usually with $\Gamma\approx 1$. This also rules out an association with background AGNs. 

Several figures throughout this paper display background-subtracted images, for visual clarity. All the data analysis, including the proper motion fits, was done on the original, non-background-subtracted data. The backgrounds used for images of a small part of the jet or counterjet were selected to be squares at a similar distance from the nucleus, of comparable size as the selected region, without containing any point-like source identified by \texttt{wavdetect}. Figures displaying the entire $2.87'\times2.46'$ image, or the entire jet and counterjet, used a square background from the southeast region, as this did not contain any detectable structure (see Fig. \ref{fig:RGB}). We selected it to be as large as possible, without containing any point-like source identified by \texttt{wavdetect}.

The code we used to generate the images and align them can be found at \citet{Bogensberger_Calculating_proper_motion_2024}. It contains three distinct codes, for the different methods we used to align the images, as described above. 

\section{Candidate jet knots} \label{sec:newknots}

We investigated variation in the jet throughout the 22 years of \emph{Chandra} observations we considered in this paper. Close to the nucleus, and along the jet axis, we detected two brighter features, which we labeled as C1 and C2 in Fig. \ref{fig:jetknotlabel}. The location of these sources raises the possibility that these are the closest jet knots to the nucleus. However, the source type of C1 and C2 has not been investigated in previous works \citep{2007ApJ...670L..81H, 2010ApJ...708..675G, 2019ApJ...871..248S}. Both sources are located at a transverse distance of about $104~\mathrm{pc}$ from the nucleus. C1 is located along the jet axis, at an angle of $54.8\degree$ to the east of the vertical axis. Its neighboring source, C2, is $\approx 20~\mathrm{pc}$ away, in a direction almost perpendicular to the jet axis. The angle from the nucleus to C2 is inclined at $65.2\degree$ relative to the vertical. This is a larger angle than any jet knot has, as these have angles relative to the vertical of between $47.9\degree$ and $60.0\degree$. This potentially excludes C2 from being a jet knot. Instead, C2 might not be related to the jet and may be an X-ray binary (XRB).

In the counterjet direction, there is also a point-like source close to the nucleus, which we label as C3. However, the angle of this source relative to the nucleus is $-146.4\degree$, which is even further away from the counterjet axis than C2 is from the jet axis. Counterjet knots are located at angles of between $-123.8\degree$ and $-127.3\degree$ to the vertical. The large angle offset between C3 and the counterjet axis makes it unlikely that it is a counterjet knot.

There is one more point-like source at a similar distance from the nucleus as C1, C2, and C3. It is labeled as X1 in Fig. \ref{fig:jetknotlabel}. We used the CIAO \texttt{wavdetect} functionality to identify other point-like sources near the nucleus but only found sources at greater distances. All four of these sources have a PSF which is consistent with that of a point-like source. Therefore, their extent cannot be used to distinguish between a jet knot or XRB interpretation.

Out of the three point-like sources close to either the jet or counterjet axis, only C1 lies within the range of angles spanned by known jet, or counterjet knots. Assuming that all four of these sources are XRBs, and assuming an isotropic distribution, the likelihood that one of them lies within either the spread of angles of the jet or counterjet, is $15\%$.

However, the assumption that the jet boundary in the image plane can be described by a triangle may be inaccurate close to the nucleus. For the following, we define the center of the jet and counterjet axis to be $54.2\degree$, and $-125.8\degree$, respectively, based on the range of angles of the previously known jet and counterjet knots. C1 and C2 are both located within $11\degree$ of this. The likelihood of two point-like sources to be located within $11\degree$ of the center of the jet, or counterjet axis, is $8.8\%$. This is unlikely, but not sufficiently unlikely to rule out the possibility of a chance alignment of two XRBs with the jet axis. 

We also investigated the spectra of these four point-like sources to further attempt to classify them as either XRBs or jet knots. We merged the spectra of these sources from all observations listed in Table \ref{tab:Cobs}. All four spectra are contaminated by a strong background due to their location, which exceeds the source signal above $3~\mathrm{keV}$. We defined the background as a circle of the same size used for source extraction, at the same radius from the nucleus, but at a different angle, with no point-like source in its vicinity. All four spectra have a limited number of counts, which limits the ability to constrain their spectral parameters and distinguish them. We fit the spectra using the XSPEC \citep{1996ASPC..101...17A} model \texttt{tbabs*powerlaw}. There were insufficient counts to fit more complex models. We used XSPEC version 12.14.0h and fitted by minimizing the C-statistic \citep{1979ApJ...228..939C}.

The spectra of C1 and C2 are more similar to those of other jet knots, such as AX1A and AX1C. They are fitted with a small degree of absorption ($N_{\rm H} = 0.46\pm0.11\times10^{22}~\mathrm{cm}^{-2}$ for C1, and $N_{\rm H} = 0.88\pm0.17\times10^{22}~\mathrm{cm}^{-2}$ for C2), and a moderately steep power law slope ($\Gamma = 1.7\pm0.3$ for C1, and $\Gamma = 2.1\pm0.3$ for C2). However, this power law slope is steeper than that of many other jet knots \citep{2010ApJ...708..675G}. In contrast, the spectra of both C3 and X1 are harder, with a greater degree of absorption ($N_{\rm H} = 4.7^{+2.2}_{-2.5}\times10^{22}~\mathrm{cm}^{-2}$ for C3, and $N_{\rm H} = 2.0\pm1.2\times10^{22}~\mathrm{cm}^{-2}$ for X1), and a poorly constrained power law slope ($\Gamma = 0.2^{+2.1}_{-1.6}\pm0.3$ for C3, and $\Gamma = 1.3\pm1.2$ for X1). The fluxes of all four of these sources are both within the range of fluxes of XRBs and jet knots. There is insufficient data to distinguish how much of the spectra of C1 and C2 is affected by their location within, or close to the jet. Although the spectra of C1 and C2 slightly favor a jet knot interpretation, there is also insufficient data for this to be a statistically significant conclusion.

The sources C1 and C2 were not detected in previous radio observations. This part of the jet appears as a continuous diffuse, and collimated stream \citep{2003ApJ...593..169H, 2010ApJ...708..675G}. 

\begin{figure}[h]
\resizebox{\hsize}{!}
{\includegraphics{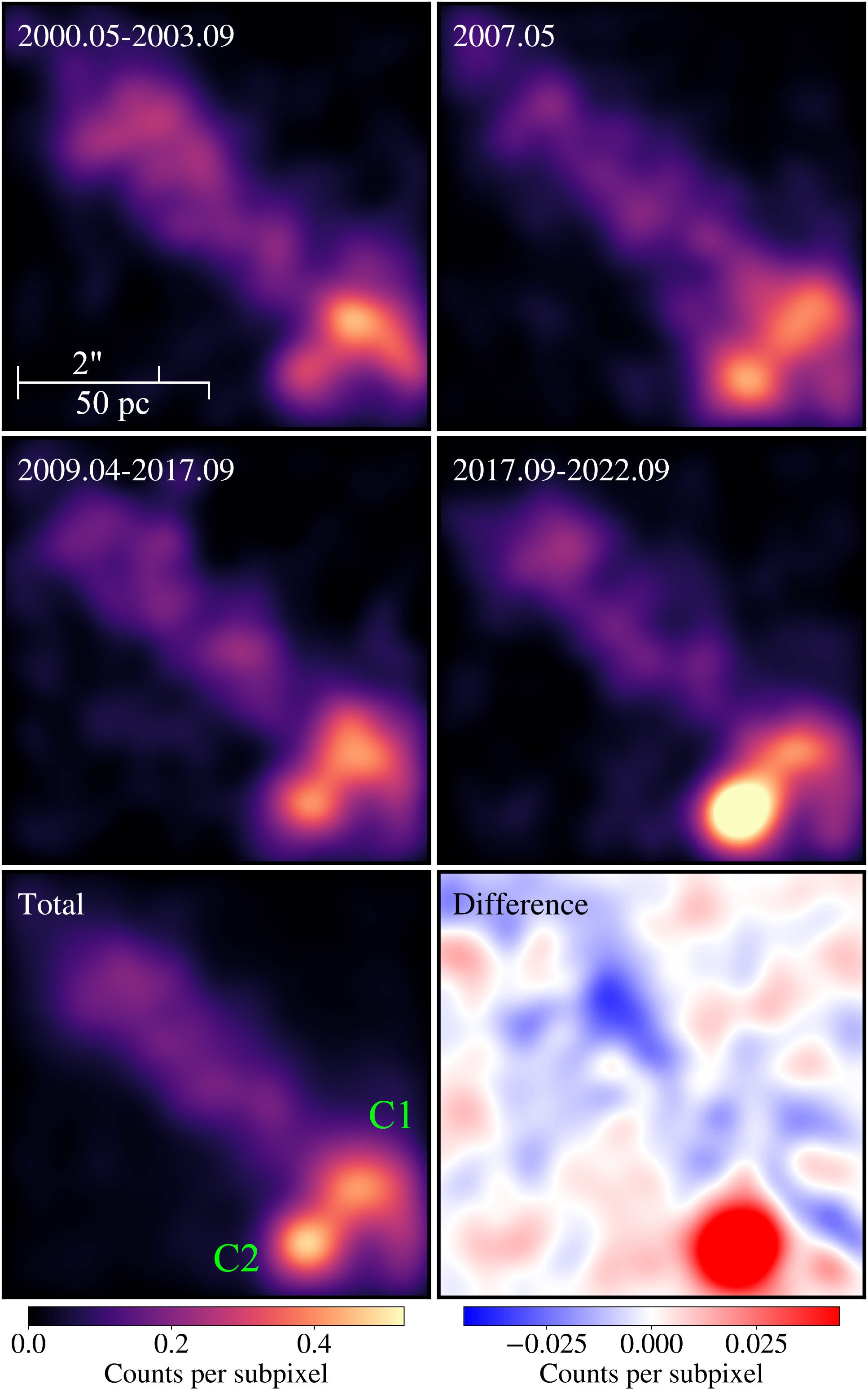}}
\caption{Gaussian-smoothed, background-subtracted images of groups of observations, and the difference image of C1, C2, and the jet region between them and the jet knots AX1A and AX1C. The colors were chosen relative to the background brightness. The difference image is calculated between the first and last groups, scaled by the number of background-subtracted source counts in these two images. Blue regions indicate parts of the image that were brighter in the first group, and red regions are parts that were brighter in the fourth group. All images depict the energy band $0.8-3.0~\mathrm{keV}$, and have a field of view of $6.0"\times6.0"$, which corresponds to $110~\mathrm{pc}\times110~\mathrm{pc}$. A colorbar is provided for the total, and difference images. The colorbars for the other panels are adjusted by the number of background-subtracted counts in this region. C1 has coordinates of $\mathrm{RA}=13\mathrm{h}\, 25\mathrm{m}\, 28.0\mathrm{s}$, $\mathrm{Dec}=-43\degree\, 01'\, 05.7 ''$. C2 has coordinates of $\mathrm{RA}=13\mathrm{h}\, 25\mathrm{m}\, 28.1\mathrm{s}$, $\mathrm{Dec}=-43\degree\, 01'\, 06.7 ''$.
 \label{fig:jkc1c2jl}}
\end{figure}

The evolution in brightness of the two sources can be seen in Fig. \ref{fig:jkc1c2jl}, which displays merged images from the four different groups of observations. Whereas C2 became brighter during these 22 years, C1 appeared to become fainter. This is of interest, as the jet knot HST-1 in M87, located at a similar transverse distance from the nucleus, was also observed to vary in brightness on similar timescales \citep{2009ApJ...699..305H}. A major difference is that HST-1 was observed to travel with a superluminal proper motion of $\approx 6$ \citep{1999ApJ...520..621B, 2019ApJ...871..248S}, whereas there is no clear evidence of significant proper motion of C1 and C2. If they are jet knots, the timescale of the variation seen in them may be attributable to impulsive particle acceleration \citep{2010ApJ...708..675G}.  

Figure \ref{fig:jkc1c2jl} also displays the diffuse X-ray emission in the region between C1, C2, and the jet knots AX1A and AX1C. This X-ray emission cannot be constrained to specific jet knots and is significantly fainter than C1 and C2. Assuming Poisson statistics, the exposure-corrected, background-subtracted count rate in the $0.5-7~\mathrm{keV}$ band remained consistent for the first three groups of observations. However, between groups 3 and 4, it dropped from $1.27\pm0.10\times10^{-5}~\mathrm{cts}~\mathrm{cm}^{-2}~\mathrm{s}^{-1}$ to $0.54\pm0.08\times10^{-5}~\mathrm{cts}~\mathrm{cm}^{-2}~\mathrm{s}^{-1}$. However, at least part of this drop is likely to be due to the decrease of the \emph{Chandra} effective area at low energies. 

\begin{figure}[h]
\resizebox{\hsize}{!}
{\includegraphics{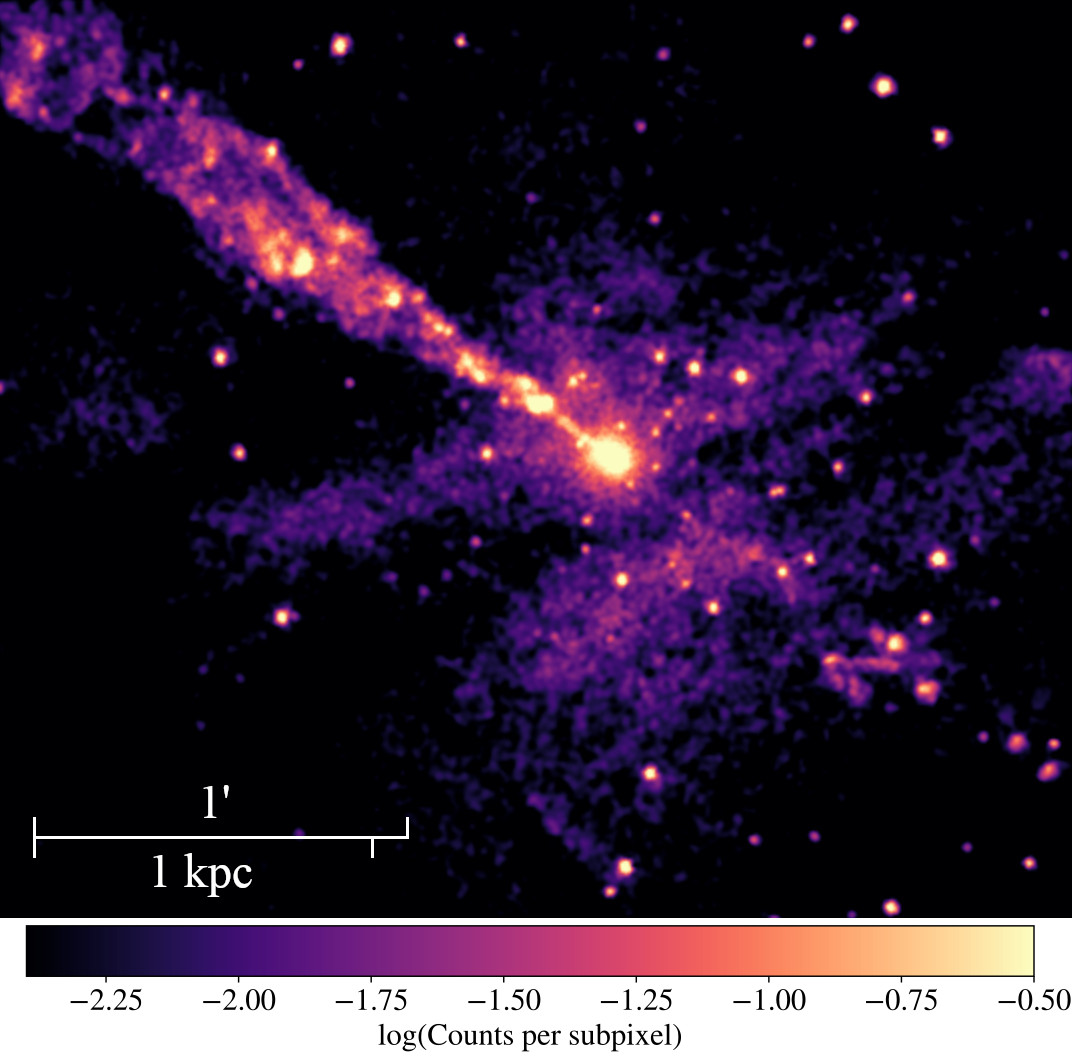}}
\caption{Adaptively smoothed, logarithmically scaled, and background-subtracted image of the $0.8-3.0~\mathrm{keV}$ energy band of the combined \emph{Chandra} observations of Cen A used in this work. Other properties are detailed in the caption of Fig. \ref{fig:RGB} 
 \label{fig:0.8-3}}
\end{figure}

We selected the energy band $0.8-3.0~\mathrm{keV}$ for Fig. \ref{fig:jkc1c2jl}. Compared to other sources of X-ray emission, the jet is brightest in this energy range, as can be seen in Fig. \ref{fig:0.8-3}. At lower energies, more of the detected source counts originate from the diffuse X-ray background. At higher energies, the nuclear X-ray emission dominates and has a wider PSF that envelops the region shown in Fig. \ref{fig:jkc1c2jl}.

Despite the challenges at higher energies, we also investigated variations in this part of the jet in the $2-10~\mathrm{keV}$ band. The effective area in this band has been largely unaffected throughout the \emph{Chandra} lifetime so far. Therefore, this band allows us to determine whether the variations seen in the $0.8-3.0~\mathrm{keV}$ range are due to changes in the jet, or rather due to the change in effective area. The weak jet emission between C1, C2, and jet knots AX1A and AX1C cannot be well resolved from the background in the $2-10~\mathrm{keV}$ band. However, C1 and C2 are resolved, and show the same trend in brightness that was found for the $0.8-3.0~\mathrm{keV}$ band. The X-ray source C1 became fainter, while C2 became brighter over the course of the observations. 

Fig. \ref{fig:RGB} shows that the X-ray jet emission is harder closer to the nucleus, but becomes increasingly soft at greater distances. This might be due to the X-ray emission becoming less energetic at greater distances, or it might be due to a decreasing obscuration. Both the jet and counterjet can be distinguished from the background, and constitute a narrow beam that widens at greater distances from the nucleus. If the outer boundaries of the distant X-ray jet cone are extrapolated back towards the nucleus, they intersect shortly upstream of the jet knots AX1A and AX1C. Excluding sources C1 and C2, the jet is more collimated closer to the nucleus. The larger angular width of the jet at larger distances might be due to a weakening collimation as the jet slows down. It may, alternatively, imply a precession of the jet axis over time. This is also supported by the existence of groups of jet knots located either on the top or bottom of the jet sheath, next to other jet knots at similar distances from the nucleus. 

\begin{figure}[h]
\resizebox{\hsize}{!}{\includegraphics{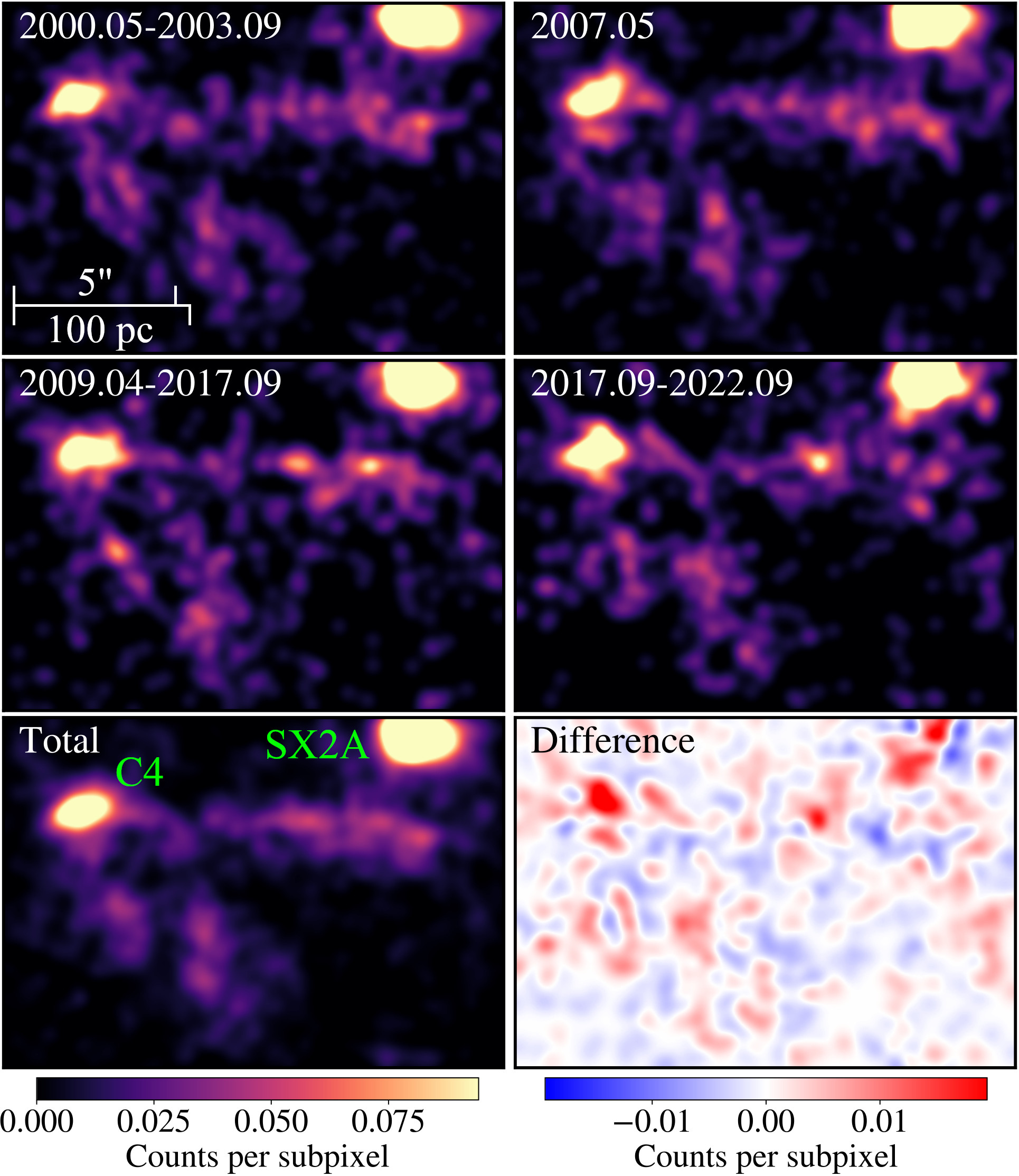}}
\caption{Gaussian smoothed, background-subtracted images of the four groups of observations of C4. The field of view of these images is $15.4"\times10.8"$, which corresponds to $283~\mathrm{pc}\times198~\mathrm{pc}$. C4 has coordinates of $\mathrm{RA}=13\mathrm{h}\, 25\mathrm{m}\, 24.4\mathrm{s}$, $\mathrm{Dec}=-43\degree\, 01'\, 41.1 ''$. Other relevant details are described in Fig. \ref{fig:jkc1c2jl}. 
 \label{fig:C4}}
\end{figure}

Near the counterjet axis, we detected a source with an unusual morphology (see Fig. \ref{fig:C4}). We label it as C4, similar to the naming of C1, C2, and C3, as the fourth candidate jet knot we are investigating here. It appears to have two streams of matter trailing away from it at two distinct angles, forming a `V'-like shape behind it. The angle between these two streams is $\approx 50\degree$, but the lower one appears to bend towards the west at larger distances. 

This source also has a significant offset from the counterjet axis, which presents a challenge for its interpretation. It has an angle of $-132.9\degree$ from the vertical, as opposed to the mean angle of the counterjet knots, of $-127.7\degree$. If the counterjet is wider than expected, then this feature could be described by the shock of jet material interacting with a non-relativistic obstacle in its path. This difference of angles is still less than those spanned by the jet knots, which range from $47.9\degree$ to $60.0\degree$. Assuming perfect symmetry, this range of angles would correspond to the interval of $-132.1\degree$ to $-120\degree$ for the angular extent of the counterjet. 

If C4 is still within the width of the counterjet, then a few other point-like sources at comparable angular offsets from the counterjet axis might also be knots. A few of these sources have, however, shown significant variability throughout the observations, which may indicate that they are X-ray binaries instead. One example of this is the point-like source slightly north of SX3B (see Fig. \ref{fig:cjetknotlabel}). 

We also investigated the spectrum of C4 and found it to be more similar to that of other jet, and counterjet knots, than to XRB spectra. It was fitted by an absorbed power law with a low $N_{\rm H}$ of $0.20\pm0.09\times10^{22}~\mathrm{cm}^{-2}$, and a photon index of $\Gamma = 1.8\pm0.3$. However, a limited number of counts again prevented the possibility of C4 being an XRB to be ruled out.

The stream of matter trailing behind C4 in the southwestern direction approximately follows the line connecting C4 with the nucleus. This kind of feature could be expected for the interaction of the jet with a high-mass-loss-rate obstacle in its path. This could produce turbulence increasing the density in the jet medium and causing an elevated X-ray emission. However, the origin of the second stream is unclear. 

Comparing the four different grouped images in Fig. \ref{fig:C4}, we find that some elements of the two streams behind C4 appear to gradually move, become brighter, or fainter. However, there are too few counts in these features to accurately identify distinct components between different observations. The difference image in Fig. \ref{fig:C4} does not feature any clearly identifiable trend in the two streams. 

\section{Proper motion of individual jet knots} \label{sec:jetvelocity}

\subsection{Methodology}\label{sec:jetvelmethod}

We investigated the possibility of measuring the proper motion of individual jet knots by tracing the movement of their count distribution through the image plane, relative to stationary sources. Instead of cross-correlating two images, or groups of images, we sought to use all available data, with the exact time of the observation. 

For this purpose, we developed a Python script to simultaneously fit all imaging data around a selected source, with a function describing the shape of a knot, which moves linearly in time within the imaging plane. The count distribution of a selected jet knot was modelled using a two-dimensional Gaussian function, with axes tilted to match the direction of the jet, or counterjet. We have made the code we used for calculating proper motions available at \citet{Bogensberger_Calculating_proper_motion_2024}.


For each jet knot, we define one axis to be parallel to the line from the nucleus to its center in the merged image of all observations. We refer to this as the parallel axis or the radial direction. We also define a second axis perpendicular to it, in a clockwise direction. We refer to this as the perpendicular axis or the non-radial direction. It points towards the northwest for the jet knots, and the southeast for the counterjet knots. 

The standard deviations of the fitting function in these two orthogonal directions were allowed to differ, to be able to better fit irregular, non-point-like shapes. We also added a fixed background to the model, which does not have any gradients within the selected region. The only component of the model that was allowed to vary in time was the position of the center of the Gaussian function, which was set to travel with a fixed proper motion along the parallel and perpendicular axes. 

Individual observations have different exposure times, vignetting strengths, and effective areas. In some observations, some jet knots are outside of the field of view. To account for this variation, we scaled the entire model of a knot by the number of counts in the selected region. The function we used can, therefore, be described as: 

\begin{align}\label{eq:velfit}
    \begin{split}
    F(x, y, t_i) &= \left(\frac{A}{2\pi\sigma_{\parallel}\sigma_{\perp}} e^{-\frac{(\Delta y'(t_i))^2}{2\sigma_{\parallel}}-\frac{(\Delta x'(t_i))^2}{2\sigma_{\perp}}}+B\right) \frac{C_i}{\sum_i{C_i}} \\
    \Delta y'(t_i) &= (y-\mu_{y})\cos{\theta}-(x-\mu_{x})\sin{\theta} - \alpha \beta_{\mathrm{app},\parallel}t_i \\
    \Delta x'(t_i) &= (x-\mu_{x})\cos{\theta}+(y-\mu_{y})\sin{\theta} - \alpha \beta_{\mathrm{app},\perp}t_i.
    \end{split}
\end{align}

\begin{figure}[h]
\resizebox{\hsize}{!}{\includegraphics{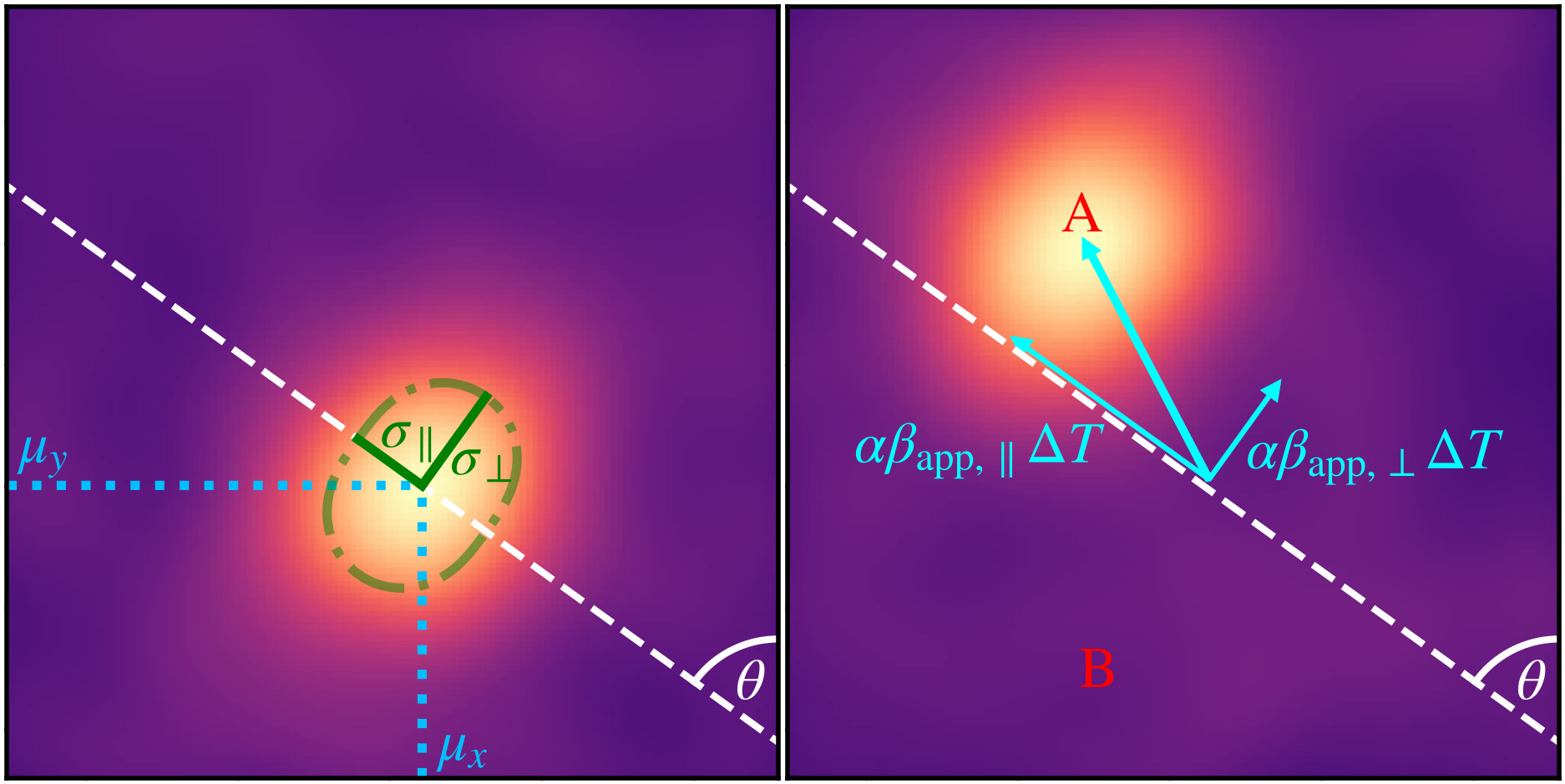}}
\caption{A visual representation of the proper motion fitting function, Eq. \ref{eq:velfit}, applied to a simulated jet knot. In this simulation, only two observations are represented here, separated in time by $\Delta T$. The displayed images are Gaussian-smoothed, akin to those of actual jet knots. $A$ represents the amplitude of the Gaussian function fitting the simulated jet knot, and $B$ is the background count rate per pixel. 
 \label{fig:PropMotExplan}}
\end{figure}

\noindent
In this equation, $F$ describes the three-dimensional fitting function, and $x$ and $y$ are the image coordinates, whereas $x'$ and $y'$ are the coordinates along the perpendicular and parallel axes, respectively. These are rotated relative to the image plane coordinates, by an angle $\theta$ in an anticlockwise direction. The parameters $\mu_{x}$ and $\mu_{y}$ describe the center of the function at time $t=0$, which we defined to occur halfway between the first and last observations, to minimize the degeneracies between the different fitting parameters. The $\sigma_{\parallel}$ and $\sigma_{\perp}$ are the standard deviations of the Gaussian in these two directions. The parameter $A$ is the amplitude of the Gaussian, describing the average number of counts in the knot, per observation. The average number of background counts in the selected region around the jet knot per observation is denoted as $B$. The parameter $C_i$ describes the total number of counts in the selected region during observation $i$, occurring at time $t_i$. The measured proper motions along the parallel and perpendicular axes are denoted as $\beta_{\mathrm{app},\parallel}$, and $\beta_{\mathrm{app},\perp}$, respectively. Finally, $\alpha$ is a constant scaling parameter that converts the proper motion into units of number of subpixels per unit time used by the parameter $t_i$. The meaning of these parameters is further illustrated in Fig. \ref{fig:PropMotExplan}.

We selected a square region, centered around a particular source, and merged the information from each subpixel of the aligned images across all observations, into a one-dimensional array, retaining the information about the $x$, $y$, and $t_i$ coordinates of each point. When fitting for the proper motions, we selected a region size of approximately twice the diameter of the selected knot, unless that included other sources, which could affect the fitting procedure. To speed up the calculation, we rebinned each image to form subpixels of 1/4 pixel size. This still benefited from the alignment of images at the 1/16 pixel size, but the results of the fits were almost identical to those found when retaining the original 1/16 pixel size. When fitting in this way, $\alpha=0.1353$, for a native pixel size of $0.492"$, and for an assumed exact distance of $3.8~\mathrm{Mpc}$ \citep{2010PASA...27..457H}. We fitted this data set using Eq. \ref{eq:velfit} and the \texttt{scipy.optimize.curve\_fit} functionality, with free parameters of $A$, $B$, $\sigma_{\parallel}$, $\sigma_{\perp}$, $\mu_{x}$, $\mu_{y}$, and most importantly, $\beta_{\mathrm{app},\parallel}$, and $\beta_{\mathrm{app},\perp}$. 

We shifted the images according to the best fit proper motion, and to ensure that $\mu_{x}$ and $\mu_{y}$ remained at the center of the selected region, then repeated the fitting procedure. It often took several iterations of this for a stable solution to be found. In the following, we always used the results of the 21\textsuperscript{st} iteration. We chose this number, as it was sufficient for a stable best fit to be found, and for us to confirm the stability of the best fit. A larger number of iterations would have yielded almost identical results. However, fitting with fewer than 10 iterations would have resulted in significantly different results in a few cases.

\subsection{Error evaluation}\label{sec:errorpropmot}

When fitting the data, we assumed that the amplitude of the jet knot profile compared to the background, and the standard deviations of the Gaussian function in the two orthogonal directions in Eq. \ref{eq:velfit} remained consistent throughout all observations. This is a simplifying assumption, which may not be correct. However, we cannot allow the widths of the Gaussian function to vary in each observation, as that hinders the ability to fit the proper motions, particularly in observations with few counts in the selected region of the jet. We investigated whether this assumption could impact our ability to constrain the proper motions of the jet knots, by simulating images of jet knots with varying widths and fluxes. We found that the proper motion could still be determined accurately in such instances, as it depends on the center of the Gaussian function, rather than on its size. However, a change in the width by more than $\approx 50\%$ sometimes caused the magnitude of the proper motion in that respective direction to be underestimated by the fits. 

We investigated to what extent our assumptions about (1) the accuracy of the image alignment, and (2) the fitting function affected the measurements of the proper motions of individual jet knots. For this purpose, we used the same methodology to fit for the proper motion of the 35 aligning sources selected in Section \ref{sec:align}, in the $0.2-10~\mathrm{keV}$ band. However, we assumed the parallel and perpendicular axes to correspond to the $y$ and $x$ axes of the image plane. The fitting function of Eq. \ref{eq:velfit} is not ideal for determining the proper motions of the aligning sources, as it assumes a constant ratio of the source to background count rate. Many of the aligning sources are strongly variable, being bright in one observation, but not detectable above background in another one. Nevertheless, the fits still constrained the proper motions of these sources. The distribution of their proper motions was centered around 0 but with a variance larger than expected from the fitted errors in the proper motion. Additionally, this deviation increased rapidly with decreasing source amplitude ($A$ in Eq. \ref{eq:velfit}). As we are assuming these aligning sources to have no proper motion, we used this excess deviation to quantify other sources of error in the proper motion measurements. Assuming an inverse relation with a constant, we found that the additional uncertainty to be best fit by:

\begin{equation}
    \sigma_{\mathrm{add}} = \frac{3.83}{A} + 0.08.
\end{equation}

We also found that the distribution of the proper motions of the aligning sources was slightly offset from 0. For those with an amplitude of $A>40 ~ \mathrm{cts}$, we found that the distribution had a mean of $-0.064$ in the $x$ direction, and $0.034$ in the $y$ direction of the imaging plane. Sources with a smaller amplitude have large uncertainties in their measured proper motions, so including them in the average would significantly distort the mean. We subtracted the components of this mean proper motion from the measurements of the jet and counterjet knots. After applying these two corrections, the errors matched the distribution of the aligning source proper motions around 0. 

This entire error estimate is biased by the use of the same sources both for the aligning of the images, and the testing of the precision of the proper motion measurements. Our choice of alignment is not perfect, but the proper motion calculation assumes it to be that way. This may introduce an additional source of error that is challenging to quantify independent of the aligning sources. For this purpose, we investigated the measured proper motions of slightly different image alignments, calculated in different ways using the different methods outlined in Section \ref{sec:align}. We considered alignments found by using a smaller sample of aligning sources, ones that were more equally distributed in the imaging plane, or only those that were located close to the jet itself. The proper motions measured for different image alignments were almost always consistent, but slightly shifted relative to each other, mostly by $<0.1$.

There is a further bias in the results, due to the selection of a Gaussian function to describe the distribution of counts in a jet knot in Eq. \ref{eq:velfit}. We repeated the calculations with a Lorentzian profile instead. The results were almost entirely consistent with those found using a Gaussian profile. We also investigated more complex functions that allowed the background to have a gradient within the image. This did not have an impact on the results of the fits, as there were mostly insufficient data to constrain a gradient in the background. Therefore, we used a constant background. 

To describe the uncertainty we have in the proper motion measurements as a result of these biases, we included an extra error, which we estimated to have a uniform value of $0.1$, for both the parallel and perpendicular directions. The three sources of error; the measurement error of fitting the data with Eq. \ref{eq:velfit}, the error found from the distribution of the proper motions of the aligning sources, and the error due to biases in the choice of alignment and fitting function, were added together in quadrature. This is the total error of the proper motion measurements, which we used below. 

We performed another check of the fitted proper motions, by limiting the analysis to only the 13 longest exposure observations. These observations are the most consistently aligned relative to each other. The proper motions found for this subset of observations were always consistent with those found for the entire set of observations listed in Table \ref{tab:Cobs}. However, the measurement uncertainties of the proper motions in this reduced sample were significantly larger. This is due to fewer total source counts, and a shorter range of observations. Therefore, we used the entire set of observations to find the best constraint on the proper motions of the jet knots. 

An additional issue is the loss of effective area at low energies over the 22 years of \emph{Chandra} observations used in this analysis. We usually used the entire \emph{Chandra} energy range, as the ability to constrain proper motions is improved by increasing the number of source counts. However, if the knots have a non-uniform spectrum across their extent, and vary in shape, then the measured proper motions may be affected by the change in effective area. We repeated the analysis to investigate if it affected our results but restricted the energy range to $2-10~\mathrm{keV}$. The effective area in this band has remained mostly consistent throughout the range of observations we used. The uncertainties for the $2-10~\mathrm{keV}$ proper motions were significantly larger, due to the reduced number of source counts available to fit with. We found that the proper motions calculated for the $0.2-10~\mathrm{keV}$ and the $2-10~\mathrm{keV}$ bands were almost always consistent within errors. In four cases the proper motion of a jet knot in one of the two directions disagreed by slightly more than $1\sigma$ in the two energy bands. Three of those were distant jet knots (CX4, FX1, and FX2), which had negligible source counts in the $2-10~\mathrm{keV}$ band, so the algorithm fitted spurious background counts instead. The fourth instance is jet knot BX4, which is surrounded by a complex jet structure, which is challenging to fit with the small number of source counts in the $2-10~\mathrm{keV}$ band.

\subsection{Results}

\begin{figure*}[t]
\resizebox{\hsize}{!}{\includegraphics{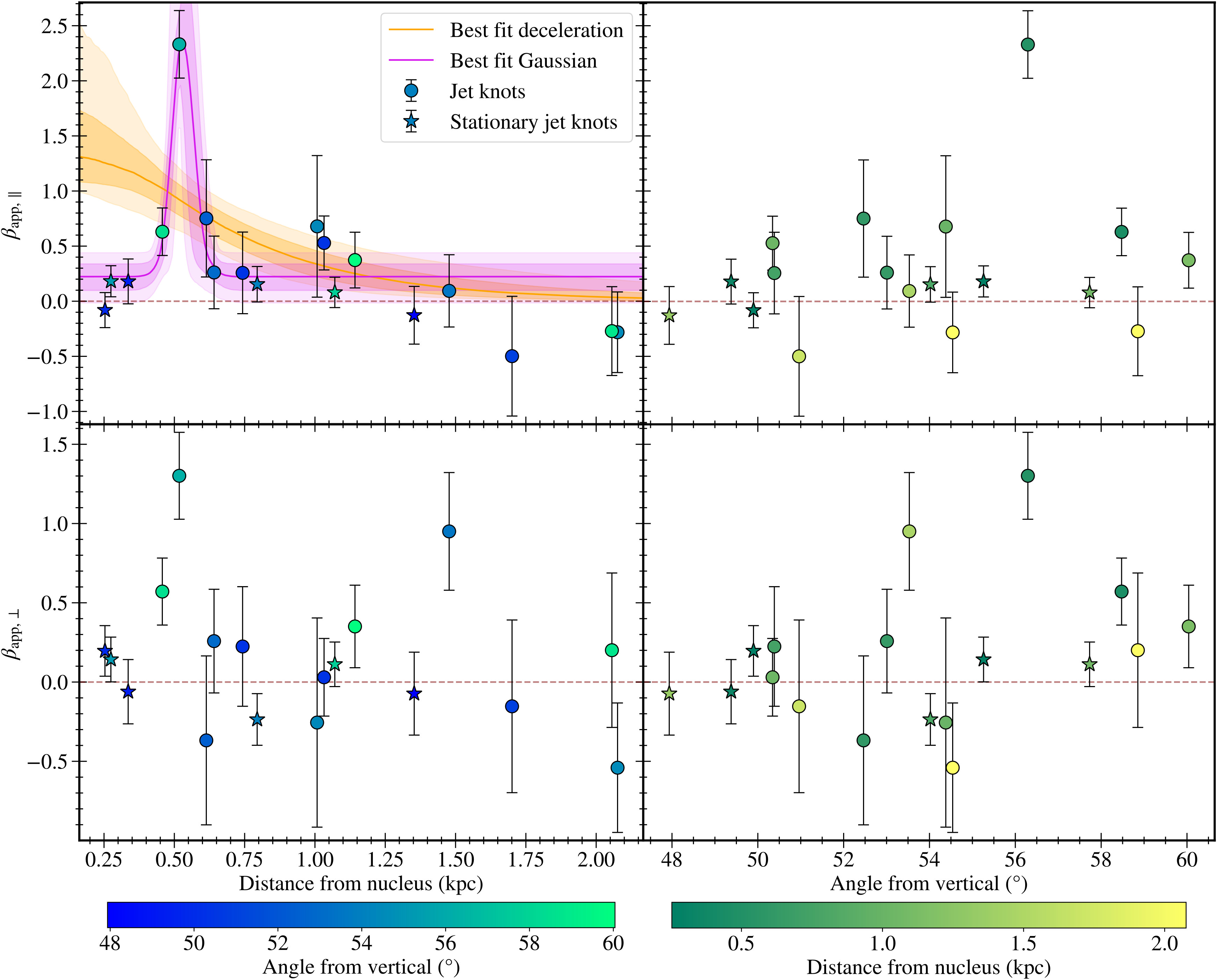}}
\caption{Apparent proper motions of the jet knots, in the parallel and perpendicular directions, as a function of distance and angle. The plots distinguish between stationary (stars), and non-stationary (circles) jet knots. Panel (a) also shows the best fit to the non-stationary jet knots. One model assumes a ram-pressure deceleration. The second model fits a Gaussian function to the data. The colored bands around the best fit depict the $1\sigma$, and $2\sigma$ confidence intervals of the best fit. 
 \label{fig:Jetvel}}
\end{figure*}

Tables \ref{tab:Jetvel} and \ref{tab:Cjetvel} list the best-fit parallel and perpendicular apparent proper motions of jet and counterjet knots. They are the results of the 21\textsuperscript{st} iteration of the fitting function described in Section \ref{sec:jetvelmethod}, corrected, and using the adapted errors found using the methodology described in Section \ref{sec:errorpropmot}. Figure \ref{fig:Jetvel} depicts the relationship between the best fit apparent proper motions of the jet knots in the parallel and perpendicular directions, as a function of the distance and angle relative to the nucleus. 

\begin{table}[h]
\centering
\setlength{\tabcolsep}{3pt}
\def\arraystretch{1.0}
\begin{tabular}{r|r|r|r|r}
    \textbf{Knot} & $\boldsymbol{D}~\boldsymbol{(\mathrm{kpc})}$ & $\boldsymbol{\theta}~\boldsymbol{(\degree)}$ & $\boldsymbol{\beta_{\mathrm{app},\parallel}}$ & $\boldsymbol{\beta_{\mathrm{app},\perp}}$ \\ \hline
    AX1A & 0.25 & 49.9 & $-0.08\pm0.16$ & $0.20\pm0.16$ \\
    AX1C & 0.27 & 55.3 & $0.18\pm0.14$ & $0.14\pm0.14$ \\
    AX2 & 0.34 & 49.4 & $0.18\pm0.20$ & $-0.06\pm0.20$ \\
    \textbf{AX3} & \textbf{0.46} & \textbf{58.5} & $\boldsymbol{0.63\pm0.22}$ & $\boldsymbol{0.57\pm0.21}$ \\
    \textbf{AX4} & \textbf{0.52} & \textbf{56.3} & $\boldsymbol{2.33\pm0.31}$ & $\boldsymbol{1.30\pm0.27}$ \\
    AX5A & 0.61 & 52.5 & $0.75\pm0.53$ & $-0.37\pm0.53$ \\
    AX5B & 0.64 & 53.0 & $0.26\pm0.33$ & $0.26\pm0.33$ \\
    AX5D & 0.74 & 50.4 & $0.26\pm0.37$ & $0.22\pm0.38$ \\
    AX6 & 0.79 & 54.0 & $0.15\pm0.16$ & $-0.24\pm0.16$ \\
    BX1 & 1.01 & 54.4 & $0.68\pm0.64$ & $-0.26\pm0.66$ \\
    \textbf{BX2A} & \textbf{1.03} & \textbf{50.3} & $\boldsymbol{0.53\pm0.24}$ & $0.03\pm0.24$ \\
    BX2 & 1.07 & 57.7 & $0.08\pm0.14$ & $0.11\pm0.14$ \\
    BX4 & 1.14 & 60.0 & $0.37\pm0.25$ & $0.35\pm0.26$ \\
    CX2 & 1.35 & 47.9 & $-0.13\pm0.26$ & $-0.07\pm0.26$ \\
    \textbf{CX4} & \textbf{1.48} & \textbf{53.5} & $0.09\pm0.33$ & $\boldsymbol{0.95\pm0.37}$ \\
    EX1 & 1.70 & 51.0 & $-0.50\pm0.54$ & $-0.15\pm0.54$ \\
    FX1 & 2.08 & 54.5 & $-0.28\pm0.37$ & $-0.54\pm0.41$ \\
    FX2 & 2.06 & 58.9 & $-0.27\pm0.40$ & $0.20\pm0.49$ \\ \hline
    
\end{tabular}
\caption{Apparent proper motions, and further properties of jet knots relative to the Cen A nucleus. The transverse distance from the nucleus to each jet knot is labeled as $D$. The angle of the line connecting the nucleus to each jet knot, relative to the vertical axis, is labeled as $\theta$, and is measured counterclockwise. The fits were performed for the $0.2-10~\mathrm{keV}$ \emph{Chandra} energy band. Rows highlighted in bold are the jet knots that have a significant proper motion of $>2\sigma$ in either the parallel or perpendicular direction. Only the proper motions that exceed this limit are highlighted in bold. 
\label{tab:Jetvel}}
\end{table}

\begin{table}[h]
\centering
\setlength{\tabcolsep}{3pt}
\def\arraystretch{1.0}
\begin{tabular}{r|r|r|r|r}
    \textbf{Knot} & $\boldsymbol{D}~\boldsymbol{(\mathrm{kpc})}$ & $\boldsymbol{\theta}~\boldsymbol{(\degree)}$ & $\boldsymbol{\beta_{\mathrm{app},\parallel}}$ & $\boldsymbol{\beta_{\mathrm{app},\perp}}$ \\ \hline
    SJX1B & 0.270 & -127.3 & $-0.48\pm0.88$ & $-0.13\pm0.88$ \\
    SX1 & 0.599 & -123.8 & $-0.10\pm0.25$ & $-0.3\pm0.25$ \\
    C4 & 0.869 & -132.9 & $0.04\pm0.46$ & $-0.60\pm0.47$ \\
    \textbf{SX2A} & \textbf{0.990} & \textbf{-123.4} & $0.01\pm0.18$ & $\boldsymbol{-0.38\pm0.18}$ \\
    SX2B & 1.137 & -126.5 & $-0.18\pm0.27$ & $0.48\pm0.26$ \\
    SX3A & 1.448 & -124.8 & $-0.42\pm0.44$ & $0.38\pm0.47$ \\
    SX3B & 1.574 & -125.5 & $-0.2\pm0.45$ & $0.40\pm0.49$ \\ \hline
    
\end{tabular}
\caption{Apparent proper motions of the counterjet knots relative to the Cen A nucleus. A more detailed explanation is provided in Table \ref{tab:Jetvel}.
\label{tab:Cjetvel}}
\end{table}

Most of the jet knots had apparent proper motions consistent with 0, in both directions. However, there are a few notable exceptions. Jet knot AX4 had the largest proper motion of $2.33\pm0.31$ parallel to the jet axis, and $1.30\pm0.27$ perpendicular to it. All proper motions are expressed as fractions of the speed of light. Jet knot AX4 will be discussed in more detail in Section \ref{sec:JKAX4}. The measured proper motions of the closest knot to it, AX3, are a lot smaller but are still the second most significant non-zero measurements. We measured proper motions of $\beta_{\mathrm{app},\parallel} = 0.63\pm0.22$, and $\beta_{\mathrm{app},\perp} = 0.57\pm0.21$ for it. The only other proper motion measurement that differed by more than $2\sigma$ from 0 was for the parallel proper motion of jet knot BX2A, of $0.53\pm0.24$, and the perpendicular motion of jet knot CX4, with a value of $0.95\pm0.37$. However, these two jet knots have proper motions consistent with 0 in the other direction. 

The four brightest jet knots; AX1A, AX1C, AX6, and BX2, have the most well-constrained proper motions. They are all consistent with being stationary within $1\sigma$ errors, or just exceeding the errors. This is also consistent with the results of \citet{2019ApJ...871..248S}. 

The measured proper motions of several other jet knots are also consistent with no motion. However, we wish to distinguish between jet knots that are consistent with being stationary due to large uncertainties caused by a limited number of source counts, and those that are stationary with a well-constrained proper motion. We noticed a dichotomy in the upper limits of the total proper motion, $\beta_{\mathrm{app}} = \sqrt{\beta_{\mathrm{app},\parallel}^2+\beta_{\mathrm{app},\perp}^2}$. Six jet knots had $1\sigma$ upper limits between $0.33-0.52$, but all other jet knots had upper limits exceeding $0.83$. Therefore, in the rest of the text, we consider these six sources, which are all also consistent with having no proper motion within $1.2\sigma$, as the "stationary" jet knots. These six jet knots include the four brightest ones, and also AX2, and CX2. Three of these are found at the smallest angles to the vertical, of all the jet knots. 

The weighted mean proper motion (weighted by the inverse square of the error) of all jet knots in the parallel, and perpendicular directions are $0.23\pm0.05$, and $0.16\pm0.06$, respectively. When excluding the six stationary sources, these averages instead increase to $0.50\pm0.09$, and $0.38\pm0.10$. However, these results are still dominated by the $>2\sigma$ proper motions of AX3, AX4, BX2A, and CX4. Without these jet knots, the weighted averages are $0.16\pm0.13$, and $0.09\pm0.14$ in the parallel, and perpendicular directions, respectively. This reflects the scatter of the jet knots detected with fewer source counts. The measured perpendicular proper motions are mainly distributed around close to 0, whereas the parallel proper are distributed around a slightly more positive value. 

Jet knot BX2 is one of the brightest jet knots, but has a proper motion consistent with 0 in both directions. It has a conical shape that differs from those of other jet knots. \citet{2003ApJ...593..169H} pointed out that it is likely to be extended, as its size and morphology are inconsistent with that of a point source. This suggests that it could be caused by the interaction of the jet with an obstacle in its path, such as a gas cloud \citep{2003ApJ...593..169H, 2010ApJ...708..675G}. As the obstacle is not moving relativistically with the jet, it does not have a measurable proper motion. The interaction of the relativistic jet with the non-relativistic obstacle would create a bow shock behind the obstacle, which could match the shape of BX2. The physical origin of the two brightest jet knots, AX1A, and AX1C has also been interpreted in this way \citet{2019ApJ...871..248S}. They also have proper motions almost consistent with 0 but have a very different shape in the image. It might also be the origin of the other jet knots we have labeled as stationary; AX2, AX6, and CX2. 

The proper motion of C4 parallel to the counterjet axis is negligible, but it is moving with an apparent perpendicular proper motion of $0.60\pm0.47$ towards the counterjet axis. If C4 is a non-relativistic obstacle in the path of the outer counterjet, whose interaction with it produces the two streaks behind it, it should not have any measurable proper motion at this sensitivity. However, the ability to measure the proper motion for this source may be affected by its unusual shape, which is very different from that assumed by Eq. \ref{eq:velfit}. Alternatively, the apparent motion may have been caused by the motion of an X-ray-emitting element along the horizontal stream behind C4. Such a structure would not be resolved but would appear to move the center of AX3 in its direction of motion. 

All counterjet knots have a parallel proper motion consistent with 0. Only one counterjet knot, SX2A, has a significant perpendicular proper motion $>2\sigma$, of $-0.38\pm0.18$. The weighted means of the proper motions of the counterjet knots in the parallel and perpendicular directions are $-0.10\pm0.12$, and $-0.08\pm0.12$, respectively. 

The proper motion of C1 and C2, the two point-like sources near the jet axis that are the closest to the nucleus, could not be determined. The strong, and variable background with a significant, and variable gradient, and the presence of other structures near these sources, which strongly deviate from the assumptions made by Eq. \ref{eq:velfit}, prevented the code from accurately tracing potential changes in their positions over time. 

\subsection{Jet knot AX4}\label{sec:JKAX4}

\begin{figure}[h]
\resizebox{\hsize}{!}{\includegraphics{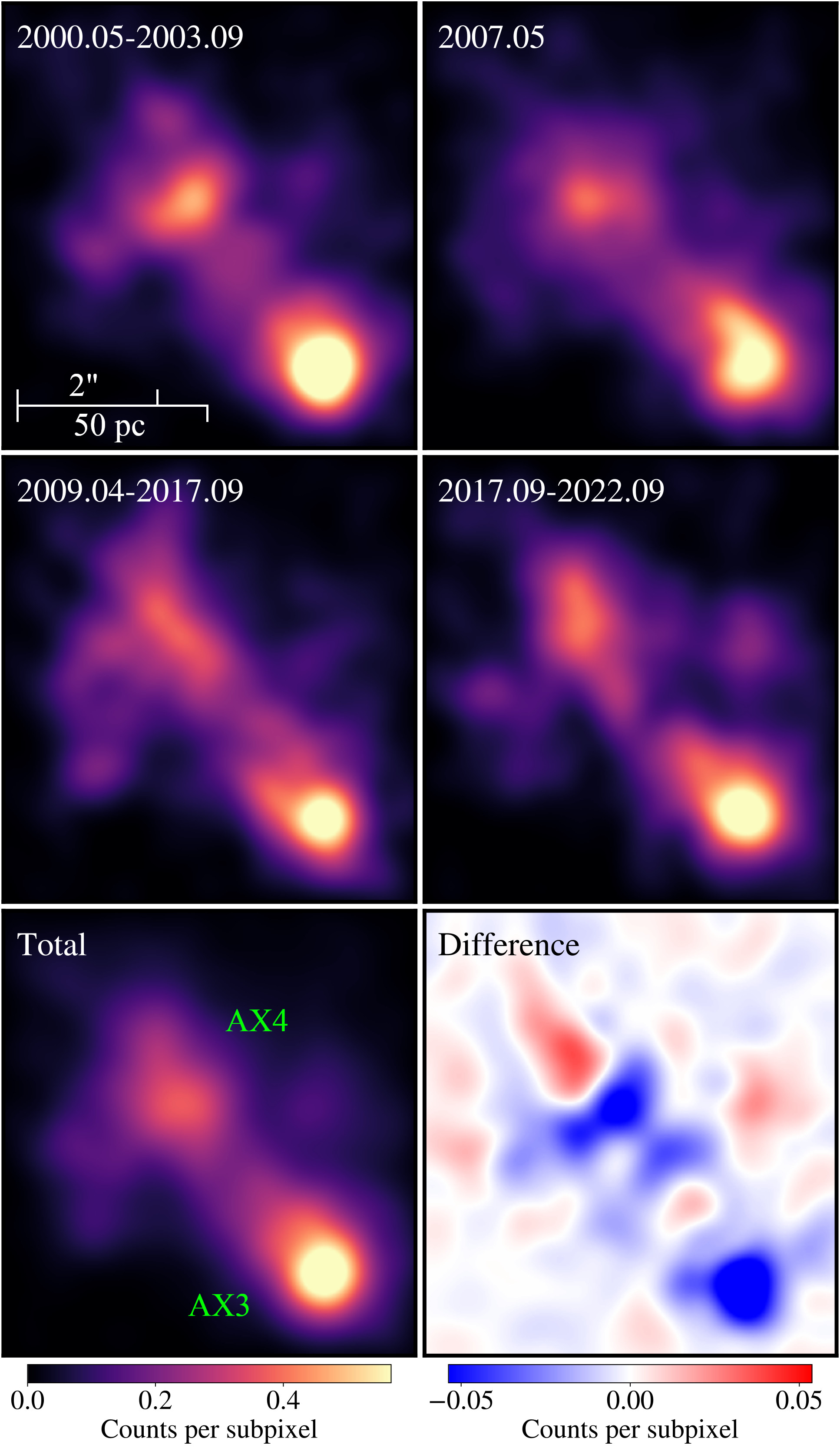}}
\caption{Gaussian smoothed, background-subtracted images of groups of observations, and the difference image of jet knots AX3 and AX4. The field of view of these images is $5.8"\times6.3"$, which corresponds to $108~\mathrm{pc}\times116~\mathrm{pc}$. AX3 has coordinates of $\mathrm{RA}=13\mathrm{h}\, 25\mathrm{m}\, 29.5\mathrm{s}$, $\mathrm{Dec}=-43\degree\, 00'\, 56.1 ''$. AX4 has coordinates of $\mathrm{RA}=13\mathrm{h}\, 25\mathrm{m}\, 29.7\mathrm{s}$, $\mathrm{Dec}=-43\degree\, 00'\, 52.9 ''$ in recent observations. Further explanation is provided in Fig. \ref{fig:jkc1c2jl}. 
 \label{fig:jkax3ax4}}
\end{figure}

Jet knot AX4 has a significantly larger measured proper motion than any other jet, or counterjet knot. It is the only knot with a measured total apparent superluminal proper motion. The evolution of both AX4 and its neighboring knot, AX3, over the course of the 22 years of \emph{Chandra} observations, is depicted in Fig. \ref{fig:jkax3ax4}.

Combining the measured motions in the parallel and perpendicular directions, we find that AX4 has a total apparent superluminal proper motion of $2.67\pm0.41$. Using Eq. \ref{eq:maxinc}, we can constrain the inclination of its motion to $i_{\mathrm{max}} = 41\pm6\degree$. Using Eq. \ref{eq:minbeta} we can also constrain its actual speed to be $\beta_{\mathrm{min}} = 0.94\pm0.02$, which corresponds to $\gamma_{\mathrm{min}} = 2.9\pm0.4$. That minimum speed of the jet knot occurs for an inclination of $21\pm3\degree$.

Regardless of the inclination of the jet, all other jet knots travel significantly slower than AX4. The jet knot AX3 has a total proper motion of $0.85\pm0.30$, which is the second largest value of $\beta_{\mathrm{app}}/\sigma_{\beta}$, of $2.81$. At an inclination of $21\degree$, that would correspond to a speed of $\beta=0.74\pm0.08$, and $\gamma = 1.5\pm0.2$. The speed of AX4 is very anomalous compared to the rest of the jet knots. 

Extrapolating its parallel proper motion back from its current position, and assuming this motion remained constant in the past, we estimate that AX4 was emitted from the nucleus $710\pm90$ years ago, as seen by an observer. However, extrapolating its perpendicular velocity back implies an origin outside of the jet, and $>290\pm70~\mathrm{pc}$ away from the nucleus. Therefore, the apparent perpendicular velocity of this knot has to have changed in the past. 

As Fig. \ref{fig:jkax3ax4} shows, AX4 also underwent a morphological variation. In the $0.8-3.0~\mathrm{keV}$ band, it was initially being wider in the perpendicular than in the parallel direction. However, in the third and fourth groups of observations, it is seen to have an elongated shape along the jet axis. This is contrasted by knot AX3, which maintained an approximately circular shape throughout all epochs. The variation in the width of AX4 is less at higher energies than in the $0.8-3.0~\mathrm{keV}$ band, but the elongation in the jet direction is seen in all energy bands. The morphological variation in AX4 further supports the measured large proper motions. For it to change as much on these timescales requires components of it to be moving with an apparent superluminal proper motion. All other jet knots had an approximately consistent morphology throughout the observations. 

In fits to simulated images of jet knots with varying morphologies, we could not obtain as large proper motions as were measured for AX4, while maintaining a stationary centroid. In simulated instances of larger morphological changes than were observed in AX4, and for a moving centroid, the proper motions were still measured accurately or were underestimated in some cases. 

The difference image, also shown in Fig. \ref{fig:jkax3ax4}, demonstrates the movement of both knots during the intervening interval. Both knots appear blue in the region they used to be located at, and red at the positions they moved to throughout the intervening interval. The shift is larger for AX4, as it has a larger proper motion. 

Another possibility to explain the anomalously large proper motion is that AX4 consists of two jet knots, one of which became significantly brighter, whereas the other became significantly fainter during this interval. However, the necessary brightness variation needed for this to happen has not been observed in any other comparable jet knot. 

To investigate this, and other possible causes of the detected large proper motion, we examined whether it remained consistent. We fitted its proper motion in observations taken in the first half of the total time span and repeated this for the second half. A consistent proper motion in the two halves, which is also consistent with that of the total interval, is expected if it represents the actual motion of the knot. If the measured long-term apparent proper motion were instead due to brightness variations of two distinct features, we would expect the apparent proper motion in at least one of the halves to be significantly slower, and inconsistent with the value for the total interval. 

The measured proper motions in the first half were: $\beta_{\mathrm{app},\parallel}=2.04\pm0.82$, and $\beta_{\mathrm{app},\perp}=0.36\pm0.57$. In the second half, the proper motions were: $\beta_{\mathrm{app},\parallel}=2.33\pm0.48$, and $\beta_{\mathrm{app},\perp}=1.38\pm0.45$. There were more observations in the second half, which allowed the proper motions to be determined more precisely for it. Both the parallel, and the total proper motions have remained at a consistently high value in both halves, both of which are also consistent with the value measured for the total interval. This indicates that the measured apparent proper motion is indeed likely to represent the actual motion of the jet knot. However, the perpendicular proper motions in the two halves differ by $1.4\sigma$, which may indicate that this component is not constant, but increased over time. 

\subsection{Correlations and trends}

\begin{figure}[h]
\resizebox{\hsize}{!}{\includegraphics{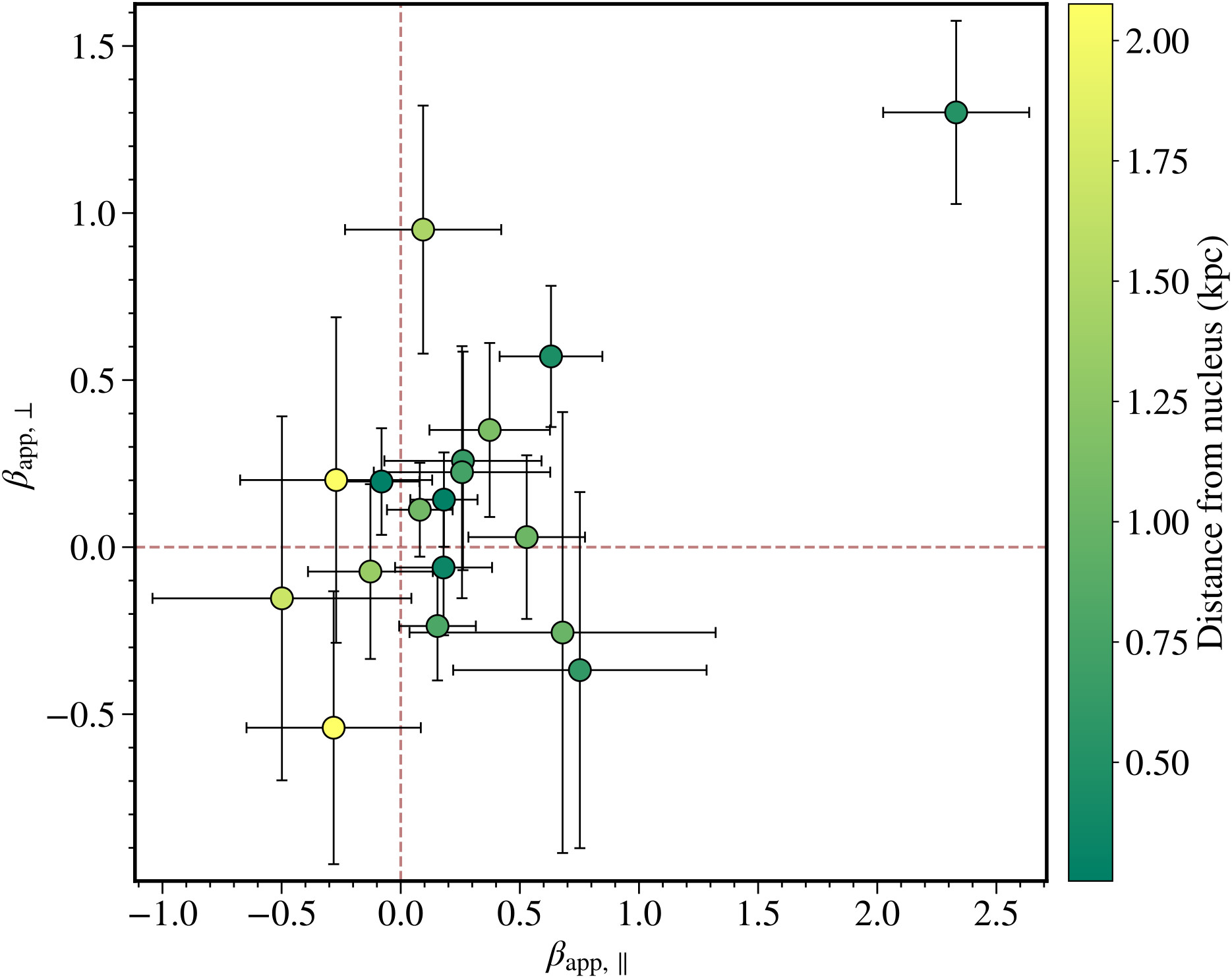}}
\caption{Apparent proper motions of the jet knots in the parallel and perpendicular directions.    
\label{fig:velparvelper}}
\end{figure}

In this section, we investigated possible correlations and trends in the measured proper motions. We used UltraNest \footnote{\href{https://johannesbuchner.github.io/UltraNest/}{https://johannesbuchner.github.io/UltraNest/}} \citep{2021JOSS....6.3001B} to calculate Bayes factors, for comparing fitted correlations against the null hypothesis of no correlation between the parameters.

Figure \ref{fig:velparvelper} compares the measured proper motions of all jet knots in the two orthogonal directions. We investigated to what extent the two parameters were linearly correlated. When including all jet knots, a linear fit was slightly preferred over the null hypothesis of no correlation, with a Bayes factor of $2.4$. However, this detected correlation is highly dependent on the measured proper motion of AX4 in both directions. Without it, the null hypothesis is strongly preferred over any linear relation. Therefore, we conclude that there is no correlation between the two parameters. 

Figure \ref{fig:Jetvel}, panel (a) shows a possible trend in the apparent proper motion along the jet axis, as a function of distance. It increases with distance until jet knot AX4. Beyond it, the proper motions gradually decrease at larger distances. It is unclear what the cause of the apparent increase in proper motion up to AX4 is. However, the decline in proper motion at greater distances could be interpreted as representing the gradual deceleration of the jet. In contrast, as Fig. \ref{fig:Jetvel}, panel (c) shows, there is no identifiable trend in the perpendicular velocities as a function of distance. 

To analyze this trend, we excluded the stationary jet knots and compared a linear decline with a flat line. The declining linear function was already strongly preferred over the null hypothesis, with a Bayesian factor of $1.4\times10^3$, and a gradient of $-0.85 \pm 0.18 ~\mathrm{kpc}^{-1}$. 

However, this linear model was potentially overly simplistic, did not capture the more complex relationship between the parallel proper motion and distance, and predicts all knots at transverse distances further than $1.6~\mathrm{kpc}$ from the nucleus to have a negative $\beta_{\mathrm{app},\parallel}$. So we also considered a slightly more complex model of jet deceleration dominated by ram pressure from interactions with the ambient medium. For this calculation, we assumed that the jet knots move at the same speed as particles in the jet, that all non-stationary jet knots have approximately the same proper motion at the same distance from the nucleus, and that other effects, which may cause an acceleration or deceleration of the jet are negligible. These assumptions are overly simplistic, but we do not have sufficient measurement precision to investigate more complex models. In this case, the instantaneous acceleration in the frame of the jet can be expressed as: 

\begin{equation} \label{eq:accel}
a'=-a_0\beta^2\gamma^2,
\end{equation}

\noindent 
where we assume $a_0$ to be a constant. 

Other models of jet deceleration describe the effect of radiation drag from an external radiation field, and disturbances of the electric potential \citep{2016MNRAS.463.3398B}, or the interaction with the disk emission \citep{1989ApJ...340..162M}. Deceleration due to mass loading from electron-positron pair production in the jet has also been investigated \citep{2017MNRAS.469.3840N}. 

The form of Eq. \ref{eq:accel} allows us to fit the relationship between $\beta_{\mathrm{app},\parallel}$ and the distance using only three free parameters: $a_0$, the inclination, and the initial speed of the jet. The best-fit result with this function is even more strongly preferred by the Bayesian likelihood ratio test, with a Bayes factor of $1.5\times 10^4$ relative to the null hypothesis. This best fit is shown in Fig. \ref{fig:Jetvel}, panel (a). The best-fit parameters are an initial speed of $\beta =0.81\pm0.09$ at a transverse distance of $460~\mathrm{pc}$ from the nucleus, an inclination of $70\pm20\degree$, and a deceleration constant of $a_0 =0.54\pm0.21~\mathrm{cm}~\mathrm{s}^{-2}$. Even though it is a significant improvement over previous models, and manages to capture the slowing decline in the apparent proper motion with increasing distance, it still does not capture the full complexity of the data. This biases, and affects the best-fit parameters, which are, therefore, likely to be less reliable than those found by other methods.   

The main challenge this model faces are the large differences between the measured proper motions of jet knots AX3, AX4, and the ones beyond it. One way to explain this is that AX4 is unusually fast, and does not follow the same relationship between speed and distance that other jet knots follow. Alternatively, AX3 could be unusually slow. This would, however, require a highly relativistic initial speed, a low inclination, and a rapid deceleration, to account for both AX4 and the slow proper motions measured beyond it. 

To showcase the discrepancy between the proper motions and a constant deceleration model, we instead fit the data with a Gaussian function plus a constant. Even though this model did not capture any declining proper motion beyond the AX5 knots, it fits the data better than any of the previously described models, with a Bayes factor of $9.4\times 10^6$. This could potentially be interpreted as a local re-acceleration of the jet, or as a single instance of an exceptionally energetic outflow from the nucleus.

When fitting $\beta_{\mathrm{app},\parallel}$ as a function of distance for all jet knots, including the stationary ones, the null hypothesis is slightly preferred over both the negative linear slope and the ram pressure deceleration model. However, the Gaussian fit to the anomalously high proper motion of AX4 is still strongly preferred over the null hypothesis, with a Bayes factor of $3.9\times10^9$. 

\begin{figure}[h]
\resizebox{\hsize}{!}{\includegraphics{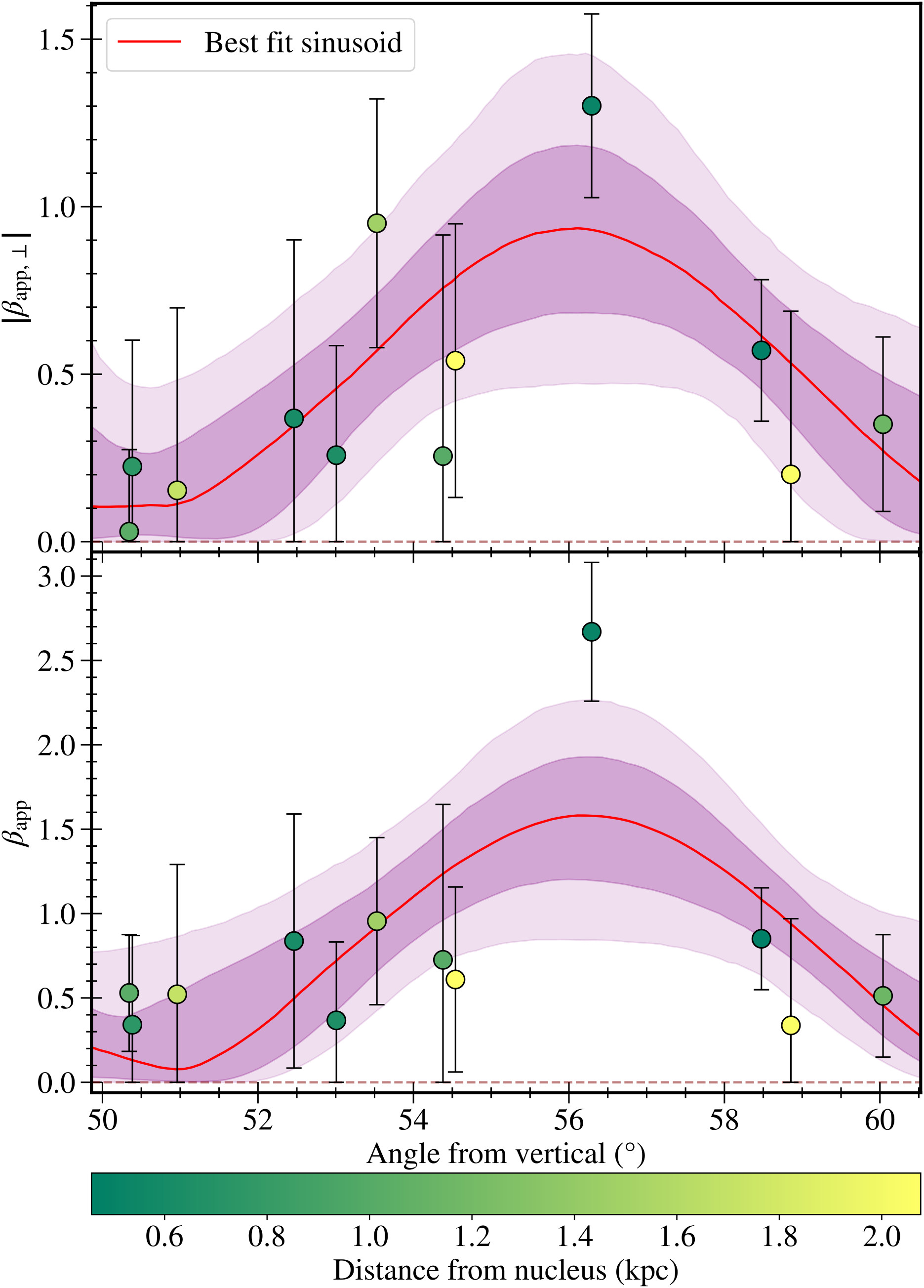}}
\caption{Modulus of the apparent perpendicular proper motion, and the total apparent proper motion, plotted as a function of angle from the vertical. The best-fit sinusoid to these data is also shown, as well as its 1, and $2\sigma$ confidence intervals. The stationary jet knots of AX1A, AX1C, AX6, and BX2 have been excluded in both panels.   
 \label{fig:PropTotvelang}}
\end{figure}

Next, we examined the dependence of the proper motions on the angle. Several jet knots exhibited perpendicular proper motions inconsistent with 0. This cannot be explained by a precession of the jet axis, as $\beta_{\mathrm{app},\parallel}$ was always measured in the direction of the line connecting the nucleus to each specific jet knot. So when extrapolating the current measured apparent proper motions of the jet knots back towards the nucleus, many jet knots would seem to not have originated from it. This suggests that the perpendicular velocities must vary over time. It must also vary on comparably short timescales, as the time needed to cross the width of the jet is often shorter than the time needed to travel from the nucleus to that point in the jet, at the currently measured proper motions. For instance, for the measured $\beta_{\mathrm{app},\perp}$ of AX4, it would take $260\pm50~\mathrm{yrs}$, as observed by a telescope, to traverse the width of the jet at its current location. In contrast, it would have taken $710\pm90~\mathrm{yrs}$ to travel from the nucleus to its location along the parallel axis, at its current $\beta_{\mathrm{app},\parallel}$.

A bending of the jet is one possible explanation for this effect. However, neither the X-ray nor the radio images show any sign of bending of the jet for the part of the jet we are investigating here. Furthermore, there is no clear trend of the perpendicular proper motion on distance from the nucleus, which would be expected if a bending of the jet were the cause of this phenomenon. 

We investigated the dependence of $\beta_{\mathrm{app},\perp}$ on the angle in the jet. This is depicted in Fig. \ref{fig:Jetvel}, panel (d). The three jet knots that have a perpendicular proper motion of $>2\sigma$ all lie towards the center, or lower left part of the jet. They all also have a positive $\beta_{\mathrm{app},\perp}$. The jet knots with the most negative values of $\beta_{\mathrm{app},\perp}$ are also found closer to the center, than at the outer sheath of the jet cone. Excluding the stationary jet knots, there appears to be a trend in which faster perpendicular proper motions in either the positive, or negative direction are found closer to the spine of the jet.

We fitted the distribution of $\left|\beta_{\mathrm{app},\perp}\right|$ as a function of the angle using: 

\begin{equation}\label{eq:circmot}
\left|\beta_{\mathrm{app},\perp}\right| = A \left(\cos\left(\frac{w\pi}{2}(\theta-\theta_0)\right)+1\right).
\end{equation}

Figure \ref{fig:PropTotvelang} plots the absolute value of $\beta_{\mathrm{app},\perp}$ as a function of the angle to the vertical, and excludes the stationary jet knots. It also displays the best fit of Eq. \ref{eq:circmot} to this data. This function was preferred over the null hypothesis of $\left|\beta_{\mathrm{app},\perp}\right|=k$, with a Bayes factor of $8.5$, and best-fit parameters of $A=0.48\pm0.12$, $w=0.32\pm0.10~\mathrm{deg}^{-1}$, and $\theta_0=56\pm1\degree$. 

One way to interpret this is to assume some helical motion of jet knots within the width of the jet. In this picture, the $\beta_{\mathrm{app},\perp}$ is a measure of the component of the circular speed in the plane perpendicular to the line of sight, after correcting for the inclination. This component should be zero at the maximum extent of the helical motion, as projected on the image plane. It should be maximally positive, or negative at angles near the center of the circular motion. This matches the distribution of $\beta_{\mathrm{app},\perp}$ as a function of $\theta$ well.  

When plotting the absolute value of $\beta_{\mathrm{app},\parallel}$ as a function of the angle to the vertical, we also found the faster jet knots to lie closer to the center (see Fig. \ref{fig:Jetvel}, panel (b)). Fitting that with Eq. \ref{eq:circmot} again prefers this function over the null hypothesis of no correlation, with a Bayes factor of 11. Faster proper motions closer to the center of the jet have also been measured in several other jets in the radio band \citep{2016A&A...587A..52M, 2017A&A...606A.103H}. 

Figure \ref{fig:PropTotvelang} also depicts the total proper motion of all jet knots, as a function of angle to the vertical. As both $\beta_{\mathrm{app},\parallel}$, and $\beta_{\mathrm{app},\perp}$ show a preference for the sinusoid fit, the same remains true for the total apparent proper motion. It is best fit with $A=0.80\pm0.18$, $w=0.32\pm0.08~\mathrm{deg}^{-1}$, and $\theta_0=56\pm1\degree$.

However, in all three cases, the properties of the best-fit sinusoid are largely dependent on the anomalously high proper motion of AX4. When excluding it, and fitting the remaining jet knots with the same two functions, the null hypothesis is preferred over the sinusoid, for all perpendicular, parallel, and total proper motions. When fitting the proper motions of all jet knots, including the stationary ones, we find the null hypothesis to be slightly preferred over the sinusoid function for the parallel, perpendicular, and total proper motions. 

\begin{figure}[h]
\resizebox{\hsize}{!}{\includegraphics{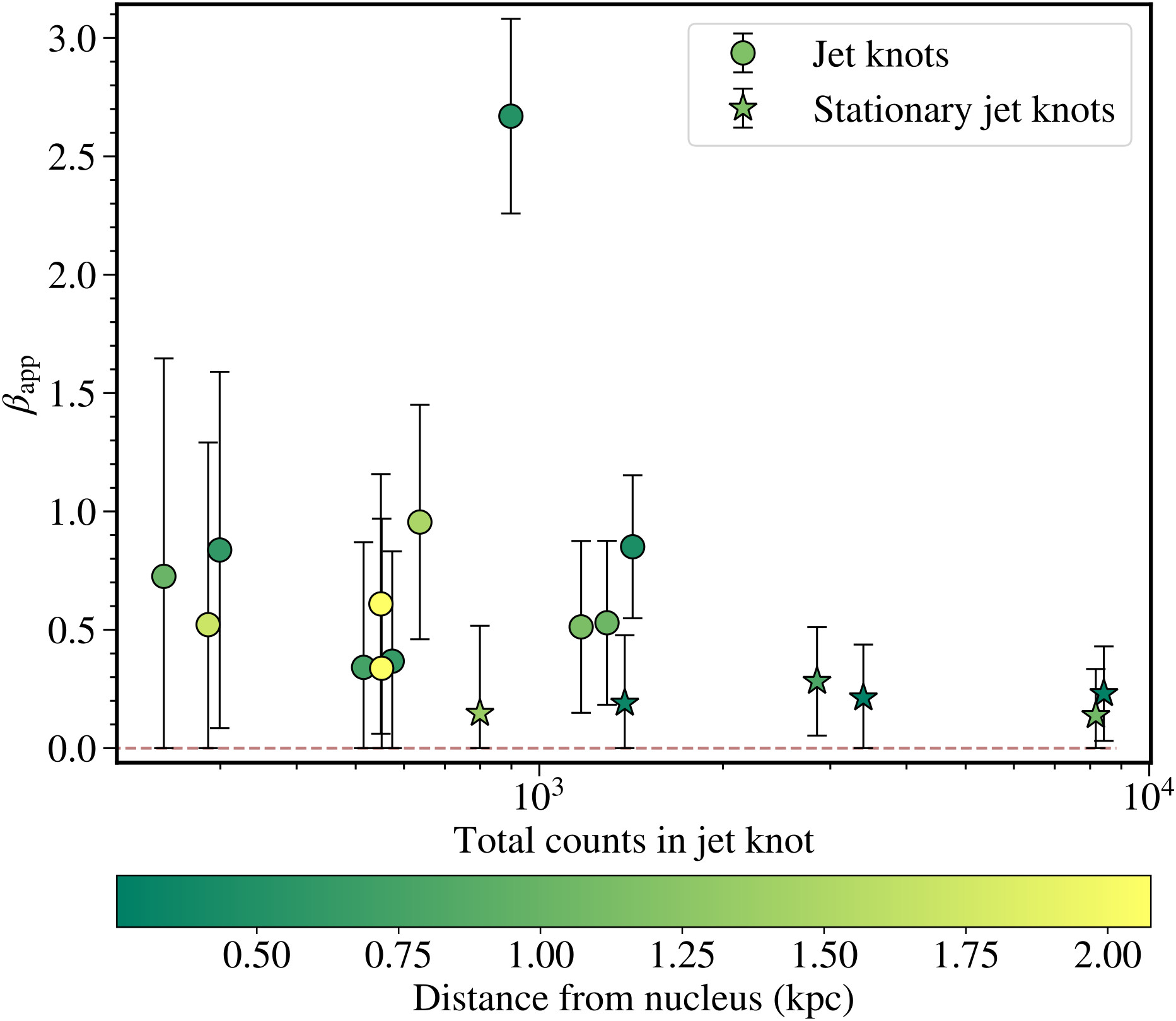}}
\caption{Total proper motion as a function of the total number of source counts.   
 \label{fig:PropMotCounts}}
\end{figure}

Finally, we investigated whether there is any correlation between the measured total proper motion with the total number of source counts of the jet knots. The dependence of these two parameters on each other is depicted in Fig. \ref{fig:PropMotCounts}. The total source counts here refer to the sum of counts fitted by the two-dimensional Gaussian function in Eq. \ref{eq:velfit}, throughout all observations.  

Brighter jet knots have smaller uncertainties, but their best-fit values do not show any clear trends with increasing source counts. When fitting the total proper motions for all jet knots, a declining power law model is barely preferred over the null hypothesis, with a Bayes factor of $1.8$. However, this is predominantly due to the brightest jet knots, all of which are stationary, and have the smallest uncertainties. When excluding the stationary jet knots, the null hypothesis is preferred. This indicates that the measurement uncertainties are reliable, and do not underestimate the level of scatter that should be expected to be found in the measured proper motions. The proper motion of AX4 is not due to the number of counts measured for it.  

\section{Discussion} \label{sec:discussion}

So far, AGN X-ray jet knot proper motions have only been measured in Cen A \citep{2019ApJ...871..248S} and M87 \citep{2019ApJ...879....8S, 2024arXiv240419272T}. The vast majority of jet proper motion measurements have been performed in the radio band, due to its better angular resolution. In these two instances, previous measurements of X-ray proper motions have agreed well with the findings in the radio band. 

In contrast, the X-ray proper motions presented here differ significantly from those in the radio band. \citet{1998AJ....115..960T, 2003ApJ...593..169H} found that the closest radio jet knots to the nucleus in Cen A were either stationary or moving with a subluminous proper motion of $\approx 0.5$. This is generally consistent with our results. However, there are significant differences when comparing the proper motions of individual jet knots in the X-ray band, with those found by \citet{2010ApJ...708..675G}. 

The X-ray knot AX4, for which we measured a total apparent proper motion of $2.67\pm0.41$, corresponds to the location of the radio knot A3B, which was found with an apparent proper motion of $0.802^{+0.15}_{-0.09}$ \citep{2010ApJ...708..675G}. The apparent X-ray proper motion disagrees with the radio measurement by $4.3\sigma$.

Furthermore, \citet{2010ApJ...708..675G} find A3B to have a significant component in the negative perpendicular direction, which also contradicts our measurement of $\beta_{\mathrm{app},\perp}=1.30\pm0.27$. Although there is a large discrepancy between these measurements, this knot was found to be the fastest-moving jet knot in both the X-ray and the radio band. The radio knot A3B is contained in a diffuse region of radio emission that extends beyond the limit of the X-ray knot \citep{2003ApJ...593..169H}. The radio and X-ray knots may correspond to different physical structures in the jet, that merely appear to be located in the same region of the jet. Beyond A3B, the radio knots A4 and B2 were found to have slightly faster median proper motions, but both were still consistent with 0 as well \citep{2010ApJ...708..675G}.  

Similarly, the radio jet knot A3A was found to have an upper limit on its apparent proper motion of $0.05$ \citep{2010ApJ...708..675G}. In contrast, the corresponding X-ray knot, AX3, was found to have an apparent proper motion of $0.85\pm0.30$. Besides A3B, the two other radio knots with identified non-zero proper motions, A1B, and A1E, have no clear X-ray counterpart. Besides AX3 and AX4, the two other X-ray jet knots with $>2\sigma$ proper motions in the parallel or perpendicular directions, BX2A, and CX4, have no identified radio counterpart.

\citet{2010ApJ...708..675G} concluded that the X-ray jet knots of Cen A correspond to stationary radio jet knots, which are shocks produced by the interaction of the jet with an obstacle in its path. Our measurement of significant X-ray proper motions of individual jet knots contradicts this. We instead demonstrate that at least four of the X-ray jet knots travel at relativistic speeds, which can significantly exceed those measured for the corresponding radio knots. Therefore, at least four X-ray jet knots are not shocked obstacles, but possibly represent blobs of compressed jet medium, or relativistic shock waves traveling through the jet medium. 

It is unclear what the cause of the large discrepancy between the radio and the X-ray measured proper motions of A3B and AX4 is. One possibility is that the diffuse radio emission around A3B hinders the ability to measure faster proper motions. Alternatively, the fast-moving X-ray component might be radio faint, as several other X-ray knots are. The association of AX4 and A3B might be wrong. It is possible that a fast X-ray-emitting component of jet material pulls some radio-emitting material along with it at slower speeds.

\citet{2019ApJ...871..248S} cross-correlated merged images of the jet taken in 2002 and 2003 (ObsIDs 02978 and 03965), with observation in 2017 (ObsIDs 19521 and 20794). The images they cross-correlated contained the jet knots AX1A to CX2, but no jet knots beyond that. They found that the peak of the cross-correlation indicated that there was no change between the two groups of images. However, when excluding the three brightest jet knots, which are also stationary (AX1A, AX1C, and BX2), and repeating the cross-correlation, they instead found a proper motion of $\beta_{\mathrm{app}}=0.68\pm0.20$. Additionally, \citet{2019ApJ...871..248S} found that the direction of the mean proper motion had a larger angle to the vertical than the mean jet axis. This can, therefore, be interpreted as a negative mean $\overline{\beta_{\mathrm{app},\perp}}$ value.

We compare these results with the measured proper motions for the individual jet knots presented in this work. These proper motions were calculated over a longer interval and used more data. The weighted average proper motions in the two directions were: $\overline{\beta_{\mathrm{app},\parallel}}=0.26\pm0.06$ and $\overline{\beta_{\mathrm{app},\perp}}=0.15\pm0.06$, with a total of $\overline{\beta_{\mathrm{app}}}=0.30\pm0.08$, when including all jet knots up to, and including CX2. When excluding jet knots AX1A, AX1C, and BX2, we find proper motions of $\overline{\beta_{\mathrm{app},\parallel}}=0.43\pm0.08$ and $\overline{\beta_{\mathrm{app},\perp}}=0.16\pm0.08$, or $\overline{\beta_{\mathrm{app}}}=0.46\pm0.11$.

The differences between these two results could be due to the inclusion of fainter, and larger, less point-like structures in the cross-correlated images. Our results are instead based only on the bright jet knots whose PSF profile can be fitted by a two-dimensional Gaussian function. This suggests that the diffuse emission which is less well constrained to a Gaussian profile might be moving with a larger apparent proper motion, and more towards higher angles, than the bright jet knots. The difference in the mean proper motion measured when including all jet knots, even the stationary ones, in the calculation, may be due to the cross-correlation focusing on the brightest features.

\citet{2019ApJ...871..248S} did not investigate the proper motions of individual jet knots. However, their difference images (their Figs. 3 and 7) do appear to show that jet knots AX3 and AX4 move significantly between the two epochs they examined. This seems to be consistent with the measured individual apparent proper motions of these two jet knots presented here, and the difference image of Fig. \ref{fig:jkax3ax4}.

The fastest jet knot proper motions are typically measured close to the nucleus after they have been accelerated for some distance \citep{2001ApJ...549..840H}. \citet{2014ApJ...781L...2A} measured the acceleration to occur over a distance of $\approx 10^6 ~ r_{\mathrm g}$ in M87, where $r_{\mathrm g}=GM/c^2$ is the gravitational radius of the black hole. \citet{2015ApJ...798..134H} found that the acceleration region can extend up to a deprojected distance of $\approx100~\mathrm{pc}$ from the nucleus. Jet knots closer to the nucleus have not yet reached the fastest speeds. More distant jet knots are found to have slower proper motions at greater distances, as they are gradually decelerated \citep{2015ApJ...798..134H}. 

In X-rays, \citet{2019ApJ...879....8S} found that the closest detectable X-ray knot to the nucleus in M87 was traveling with the largest proper motion. Jet knots at closer distances to the nucleus, which were still in the process of accelerating up to the highest speeds could not be resolved in X-rays. \citet{2019ApJ...879....8S} also found that the knot proper motion quickly declined as a function of distance for the other jet knots. 

Therefore, it is surprising that the X-ray jet knot with the largest proper motion in Cen A is located at a transverse distance of $520~\mathrm{pc}$ from the nucleus. Based on the limit on the inclination from its apparent proper motion, this corresponds to a deprojected distance from the nucleus of $>790~\mathrm{pc}$. Assuming a supermassive black hole mass of $5.5\times10^7 ~\mathrm{M}_{\odot}$ \citep{2009MNRAS.394..660C}, that is $>3\times10^8~r_{\mathrm g}$. This distance is far too large for the accelerating region. 

A smaller acceleration region implies that AX4 would have been even faster in the past, as it would have decelerated after reaching its maximal speed, and before reaching its present location. This would provide a narrower limit on the range of inclinations the jet can have than the one we found, of $i<41\pm6\degree$. This inclination constraint applies, regardless of the physical origin of the jet knot.

If all jet knots were accelerated to the same peak speed before following the same deceleration as a function of distance, AX3 should have an even faster apparent proper motion than AX4. Additionally, to explain the drop in apparent proper motion from AX4 to AX5A would require exceptionally relativistic initial speeds and significant deceleration beyond the location of AX4.  Therefore, the existence of non-stationary jet knots that have significantly slower apparent proper motions than AX4, both slightly upstream, and slightly downstream of it, suggests that X-ray jet knots are not all accelerated to the same initial speeds.

We suggest that AX4 was accelerated to far higher energies than most other X-ray jet knots and that this does not happen frequently. This would require a comparably short-term modification of the typical physical properties of the jet launching region. One possibility is that this could have been caused by a tidal disruption event, as these can produce a short term jet enhancement \citep{2020NewAR..8901538D}. 

The inclination constraint based on the apparent proper motion of AX4 is consistent with the one found on sub-pc scales, of $12–45\degree$ \citep{2014A&A...569A.115M, 2021NatAs...5.1017J}. However, it disagrees with earlier high inclinations estimates of $50-80\degree$ \citep{1994ApJ...426L..23S, 1996ApJ...466L..63J, 1998AJ....115..960T}. This also indicates that the jet is misaligned relative to the host galaxy by at least $31\pm7\degree$. 

Three jet knots and one counterjet knot also show apparent proper motion in a direction perpendicular to the jet, or counterjet axis, at a $>2\sigma$ significance. A further four jet knots and two counterjet knots have a perpendicular motion with a significance of between $1$ and $2\sigma$. For comparison, there are three jet knots with $\beta_{\mathrm{app},\parallel}>2\sigma$, and a further five with a significance between $1$ and $2\sigma$. Not a single counterjet knot has $\left|\beta_{\mathrm{app},\parallel}\right|>1\sigma$.

A non-radial motion in the Cen A jet was first proposed by \citet{2008ApJ...673L.135W} to explain the steepening of the X-ray spectrum of the jet at greater angles from the center. They suggested that these non-radial motions may be measurable in the future.

This non-radial motion has also been detected in several other jets \citep{2001ApJ...549..840H, 2009A&A...508.1205B, 2014A&A...566A..26M}. \citet{2016AJ....152...12L} found that $32\%$ of jet features with $\gtrsim3\sigma$ proper motions have a significant non-radial component. This indicates that the motion of the jet knots is not purely ballistic. 

The physical cause of this non-radial motion is not understood yet. It could be caused by helical magnetic fields \citep{2014A&A...566A..26M, 2021ApJ...923L...5P}, along which jet knots travel. Alternatively, conical shocks from the interaction of the jet with the ambient medium could also produce some non-radial motion \citet{2017MNRAS.469.4957B}. Another possibility is the impact that magnetic reconnection has on local jet motion \citep{2023ApJ...952..168M}.

We found the distribution of non-radial apparent proper motions as a function of angle is consistent with a helical motion of the knots around the jet axis. However, this fit depends on the proper motion of jet knot AX4, without which there is currently insufficient data to significantly distinguish helical motion from noise. 

There are two point-like sources (C1 and C2), which are located near the jet axis, and $104~\mathrm{pc}$ away from the nucleus. We compared their properties with two other point-like sources at a similar distance from the nucleus, which are likely to be XRBs. C1 and C2 were not previously investigated, and have not previously been classified as either jet knots or XRBs. Their location relative to the jet is moderately unlikely but could be a chance occurrence. Their spectra are more similar to those of jet knots than those of XRBs, but there are insufficient counts to significantly rule out an XRB spectrum. Further observations are required to assess the nature of C1 and C2 either as jet knots, or as XRBs.

\section{Conclusions} \label{sec:conc}

We analyzed the entire archive of previous \emph{Chandra} ACIS observations of Cen A to investigate the proper motion and variability in its jet and counterjet over more than 22 years. We excluded observations in which the jet was located far off-axis, and in which the readout streak was along the jet axis. The images were aligned by cross-correlating the images of 35 point-like sources near to nucleus. We found the region of the jet closest to the nucleus to become fainter, and investigated the increasing and decreasing brightness of two knot-like features near the base of the X-ray jet. We also detected a potential counterjet knot trailed by two streams, forming a `V'-like shape, which might be the result of jet interaction with a stationary obstacle in its path.  

Next, we jointly fitted the distribution of counts in regions around particular jet knots, across all aligned images using a two-dimensional Gaussian function that moves linearly in time over a constant background. This allowed us to measure the apparent proper motions of each of the jet, and counterjet knots. We measured these in a radial, and non-radial direction. 

The fastest jet knot, AX4, was measured to have an apparent proper motion of $2.33\pm0.31$ in the radial direction, and $1.30\pm0.27$ in the non-radial direction. Here, we express the speeds as a fraction of the speed of light. Remarkably, it is located at a distance of $520/\sin i~\mathrm{pc}$ from the nucleus, where $i$ is the inclination of the jet. There are four clearly identifiable jet knots closer to the nucleus than it, but none of them is traveling comparably fast. Three of them are likely to be stationary jet knots caused by the interaction of the jet with an obstacle in its path. 

The superluminal apparent proper motion of AX4 allows us to place a limit on the inclination of the jet, of $i<41\pm6\degree$. This limit agrees well with the measurements on sub-parsec scales, but it contradicts several previous high-inclination estimates. The minimal speed that AX4 can have to still achieve this proper motion, is $0.94\pm0.02$, at an inclination of $21\pm3\degree$. It is possible that AX4 used to travel even faster when it was closer to the nucleus, which would suggest a lower inclination.  
Besides AX4, there are two other jet knots that had $>2\sigma$ apparent proper motions in the radial direction. However, their proper motions are only a fraction of that of AX4. There is some indication of a gradual deceleration over increasing distance, but the data are best fit by assuming AX4 is traveling with an unusually relativistic speed. 

The Cen A jet in the X-ray band is notably different from that in the radio band. Several X-ray jet knots are not detected at all in the radio band, whereas other radio knots are not detected in the X-ray band. Notably, AX4 has an apparent proper motion that is $4.3\sigma$ larger than that of the corresponding radio knot A3B, which was measured to be $0.802^{+0.15}_{-0.09}$. 

Three jet knots, and one counterjet knot also featured significant non-radial proper motions. We found that the magnitude of this proper motion was typically higher towards the center of the jet.

\noindent

\section{Acknowledgements}

\noindent
This research has made use of data obtained from the \emph{Chandra} Data Archive and the \emph{Chandra} Source Catalog, and software provided by the \emph{Chandra} X-ray Center (CXC) in the application packages CIAO and Sherpa.

This work made use of the software packages astropy \citep[\href{https://www.astropy.org/}{https://www.astropy.org/};][]{2013A&A...558A..33A, 2018AJ....156..123A, 2022ApJ...935..167A}, numpy \citep[\href{https://www.numpy.org/}{https://www.numpy.org/};][]{harris2020array}, matplotlib \citep[\href{https://matplotlib.org/}{https://matplotlib.org/};][]{Hunter:2007}, scipy \citep[\href{https://scipy.org/}{https://scipy.org/};][]{2020SciPy-NMeth}, and Ultranest \citep[\href{https://johannesbuchner.github.io/UltraNest/}{https://johannesbuchner.github.io/UltraNest/};][]{2021JOSS....6.3001B}.

WNB thanks the Chandra X-ray Center grant GO2-23083X and the Penn State Eberly Endowment.

We thank the anonymous referee for their supportive comments.

\bibliography{bibliography}{}
\bibliographystyle{aasjournal}

\end{document}